\documentclass[12pt,tightenlines,eqsecnum,floats,shownopacs,nofootinbib,amsmath,amssymb,aps,prd]{revtex4}

\usepackage{color}
\usepackage{graphicx}
\usepackage{amsmath,amssymb}
\usepackage{verbatim}

\def\be{\nopagebreak[3]\begin{equation}}
\def\ee{\end{equation}}
\def\ba{\nopagebreak[3]\begin{eqnarray}}
\def\ea{\end{eqnarray}}

\def\lp{\ell_{\rm Pl}}

\def\f{\frac}

\def\rcr{\rho_{\rm max}}
\def\hom{\rm hom}

\def\t{\tilde}
\def\h{\hat}

\def\tr{\rm Trun}

\def\pphi{p_{(\phi)}}

\def\ep{\epsilon}
\def\x{\vec{x}}

\def\vk{\vec{k}}

\def\vpone{\varphi^{(1)}}

\def\pione{{\pi}^{(1)}}

\def\Q{\mathcal{Q}}

\def\qzero{\mathring{q}}
\def\Vzero{\mathring{V}}

\def\e{\mathfrak{e}}

\def\q{\mathfrak{q}}

\def\pp{\mathfrak{p}}

\def\T{\mathcal{T}}
\def\ps{\Gamma}

\def\H{\mathcal{H}}
\def\Hp{\mathcal{H}_{\rm phy}}

\def\b{{\rm b}}

\def\d{{\rm d}}
\def\q{\mathring{q}}
\def\e{\mathring{e}}
\def\V{\mathring{V}}
\def\ow{\mathring{\omega}}
\def\Tr{{\rm Tr\,}}
\def\ihalf{\f{i}{2}}
\def\p{\partial}
\def\la{\langle}
\def\ra{\rangle}

\def\l{\lambda}
\def\t{{t}}
\def\v{{\rm{V}}}

\begin{document}

\title{Loop Quantum Cosmology\footnote{Chapter contribution to ``Handbook of Space-time", edited by A. Ashtekar and V. Petkov, Springer-Verlag.}}

\author{Ivan Agullo}\affiliation{Center for Theoretical Cosmology, DAMTP, University of Cambridge, Wilberforce Road, Cambridge, CB3 OWA, U.K.}\email{I.AgulloRodenas@damtp.cam.ac.uk}

\author{Alejandro Corichi}
\affiliation{Centro de Ciencias Matem\'aticas,
Universidad Nacional Aut\'onoma de M\'exico,
UNAM-Campus Morelia, A. Postal 61-3, Morelia, Michoac\'an 58090,
Mexico}\email{corichi@matmor.unam.mx}

\begin{abstract}
This Chapter provides an up to date, pedagogical review of some of the most relevant advances in loop quantum cosmology. We review the quantization of homogeneous cosmological models,  
their singularity resolution and the formulation of effective equations that incorporate the main quantum corrections to the dynamics. We also summarize  the theory of quantized metric perturbations  propagating in those quantum backgrounds. Finally, we describe how this framework can be applied to obtain a self-consistent extension of the inflationary scenario to incorporate quantum aspects of gravity, and to explore possible phenomenological consequences.

\end{abstract}


\maketitle

\vfill
\eject

\tableofcontents

\vfill
\eject

\section{Introduction and Summary}
\label{s1}

In this volume there is an introduction to cosmology and CMB physics by Sourdeep, and on the inflationary paradigm by Wands. They summarize the synergy between theory and observations that has produced spectacular advances in our understanding of the universe in the last decades. The emergence of a ``concordance model" is a remarkable success of cosmology and the theory of General Relativity in which the current paradigm relies.  However, the widely accepted Hot Big Bang  scenario, regarded as the ``standard model of cosmology", contains important limitations, already manifest in its name. The model encompasses a phase in the very early universe in which the density of matter and the space-time curvature grow unboundedly, blowing up at the big bang singularity. The big bang is {\em not} a prediction, but the result of applying the theory {\em beyond its domain of validity}. When the energy density and curvature approaches the Planck scale, the predictions of General Relativity are unreliable; the {\em quantum} aspects of the gravitational degrees of freedom are expected to dominate in that regime. This chapter provides a possible quantum gravity extension of the well established cosmological model from the perspective of loop quantum gravity.

Loop quantum cosmology (LQC) arises from the application of  principles of loop quantum gravity (LQG) \cite{lqg} to cosmology. The goal is to quantize the {\em sector} of General Relativity containing the symmetries of cosmological space-times, by following the physical ideas and mathematical tools underlying LQG, presented in detail in the chapter by Sahlmann. Restricting attention to cosmology presents several advantages. The existence of underlying symmetries largely simplifies technical issues, and allows to overcome mathematics difficulties that are hard to handle in more generic situations. Yet, the structure is rich enough to contain deep conceptual issues in quantum gravity: What happens with space and time when matter density and curvature reach the Planck scale. Does the big bang singularity persist? What is the meaning of time in the Planck era? How do classical General Relativity and a smooth space-time description arise in the low energy regime? What is the scale at which quantum gravity effects  become subdominant? Does quantum gravity have anything to contribute to the origin of cosmic structures and to the inflationary scenario? 
On the other hand, the astonishing advances in theoretical  and observational Cosmology in the last years have been able to relate observations with theories of the very early universe. Cosmology then offers an interesting arena in which quantum gravity can make contact with other theories such as inflation, and probably provides the most promising avenue to confront quantum gravity ideas with observations.

But the restriction to cosmological settings also leads to important limitations. In principle, it is not guaranteed that the result of quantizing a symmetry reduced sector of general relativity will reproduce the same physics as the restriction of a full quantum gravity theory to symmetric scenarios. Symmetry reduction often entails a drastic simplification, and one may loose important features of the theory by restricting the symmetry prior to quantization. However, it has been extremely useful in several areas of physics, when the complexity of the problem under consideration made it difficult to find solutions without introducing additional inputs. The Oppenheimer-Snyder model of black hole formation, or the Dirac quantization of the hydrogen atom are examples that were able to encode the key physical ingredients of the problem, in spite of the severe symmetry reduction. Quantum cosmology may well be another example, if it is constructed choosing carefully the key ingredients from full quantum gravity. It is likely that predictions from quantum cosmology  will not agree in every detail with those obtained from full quantum gravity applied to cosmological scenarios, but  we expect it to capture the main aspects of the complete theory. As in the previous examples, quantum cosmology can provide valuable information about the correct way to quantize gravity, and be as useful as the hydrogen atom has been for quantum mechanics.

This chapter provides a brief and pedagogical summary of the advances  in Loop Quantum Cosmology, with some emphasis on recent results. They can be divided into three parts, which are in one-to-one correspondence to the three sections in which the chapter is divided: 1) Quantization of cosmological space-times; 2) Inhomogeneous perturbations in LQC; 3) LQC extension of the inflationary scenario. In the remainder of this introduction we summarize the content of each of these sections and provide a global picture.\\

{\bf 1) Quantization of cosmological space-times.} 
General Relativity is a totally constrained theory, in the sense that the full Hamiltonian generating dynamics is  required to vanish. Something similar happens in classical electromagnetism, where {\em part} of the Hamiltonian, the piece that generates gauge transformations, is a constraint. In General Relativity the constraint turns out to be the {\em full} Hamiltonian, reflecting the background independence of the theory. Dirac provided the conceptual framework to quantize constrained systems.   At the quantum level, physical states have to be annihilated by the operator corresponding to the classical Hamiltonian, $\hat{\cal C}\, \Psi=0$, and all the physics has to be extracted from this equation. The quantum state $\Psi$ is the wave function of the physical fields, including the gravitational field itself, and classical quantities such as the metric, energy density, curvature tensor, are represented by quantum operators on the physical Hilbert space ${\cal{H}}_{\rm phy}$ it belongs to. The non-trivial mathematical problem is to make sense and solve the quantum constraint equation, and  the underlying cosmological symmetries largely facilitate this task.

The next conceptual issue is to obtain the familiar time evolution that we normally use in physics from this time-less or `frozen' formalism. At the quantum level we do not have a classical metric telling us what are the time-like directions in the manifold, and all what we have is a probability-distribution $\Psi$ of different metrics. A useful strategy has been to follow a `relational-time' approach, in which one of the physical variables of the problem plays the role of time, and the rest evolve with respect to it. By using a {\em massless scalar field as this internal time}, it is possible to construct the Hilbert space of physical states satisfying the quantum constraints, and a precise mathematical framework has been developed to to study the resulting quantum geometry \cite{abl}. It has been shown that all the operators representing physical quantities such as the energy density, space-time curvature, etc, {\em remain bounded on the physical Hilbert space}, even in the deep Planck regime. This is the mathematical sense in which the singularity is resolved in LQC. The physical picture that emerges from the abstract formalism is the following. When the energy density of the universe is comparable to the Planck energy density, the quantum properties of space-time geometry become important and dominate. A sort of quantum repulsive degeneracy force appears at such extreme densities, precludes the universe to continue contracting, and forces the quantum space-time to expand again once the maximum energy density has been attained, replacing the big bang singularity by a {\em quantum bounce}. This maximum energy density is proportional to $\hbar^{-1}$, similar to the finite energy of the ground state of the hydrogen atom that avoids the collapse of the positron and electron as a consequence of the Heisenberg principle. When the energy density and curvature become smaller than approximately one percent of the Planck scale, the quantum effects of gravity become rapidly negligible and classical General Relativity provides an excellent approximation. The resulting quantum dynamics has been analysed in detail and has provided important insights on the behaviour of physics in the Planck regime.  The ability of incorporating non-perturbative quantum corrections that are able to completely dominate the evolution in the Planck regime and dilute the big bang singularity and, at the same time, to disappear in the low energy regime to find agreement with the classical description, is a highly non trivial result of LQC.  
 
Remarkably, some global aspects of the evolution of the quantum geometry can be encoded in simple {\em effective equation}. Those equations provide a smooth space-time metric that approximates the full quantum evolution of the quantum  space-time. They have similar form to the equations arising in General Relativity, but include new terms, proportional to $\hbar$, that make the effective trajectory to depart from the classical one around the Planck era. The effective dynamics provides an excellent approximation of the quantum evolution, even at Planckian densities, provided the quantum state is chosen to be highly peaked in a classical trajectory in the low energy regime where General Relativity provides a good approximation.\\

{\bf 2) Inhomogeneous perturbations in quantum cosmology.} As emphasized in the chapters by Sourdeep and Wands, the theory of inhomogeneous perturbations (of  matter and gravitational degrees of freedom)  propagating in classical cosmological space-times has been a key mathematical tool in modern cosmological research. One of the deepest insights in  cosmology is the idea that the cosmic structures (galaxy clusters, super-clusters, etc) that we see today were originated in the very early universe by a process of {\em amplification of quantum fluctuation by the cosmological  expansion}, as explained in the context of cosmic inflation in the chapter by Wands. In the inflationary scenario, this occurs  when the energy density in the universe was close to the GUT scale, $(10^{16} {\rm GeV})^4$, around 12 order of magnitude below the Planck energy density.  Quantum gravity effects of the background space-time metric are subdominant at those scales, and the theory of quantized fields propagating in a {\em classical} background appears to be the appropriate mathematical framework to work out physical predictions. However, earlier in the evolution of universe, when the curvature and energy density are close to the Planck scale, quantum gravity effects are expected to be important, and they should not be ignored.  To have a complete picture of the evolution of cosmic inhomogeneities that encompasses the Planck regime we need to learn how quantum fields propagate on a {\em quantum cosmological space-time} \cite{akl,aan2}. The  goal of the second section of this chapter is to review the construction of  such a theory.

The detailed description of quantum cosmologies provided by LQC is the suitable arena. The construction of QFT on quantum cosmologies follows closely the guiding principle behind LQC: first carry out a truncation of the classical theory  adapted to the given physical problem, and then quantize by using LQG techniques. The sector of the classical theory of interest is  {\em extended} in this part to cosmological background {\em plus first order inhomogeneous perturbations on it}. 

The resulting framework originates from first principles, under the assumption that inhomogeneous behave as {\em test fields} on the quantum geometry, and it should provide a bridge between quantum gravity and QFT on curved space-times. Therefore, it is suitable to face important conceptual questions such as: What are the concrete approximations under which the familiar quantum field theory (QFT) in classical space-times arises from this more complete description? What are the precise aspects of the quantum geometry that are `seen' by the quantum fields propagating on it? Does the resulting QFT make sense for trans-Planckian modes? These issues will be discussed with some detail in Section~\ref{sec:2}. In Section~\ref{sec:3}, this framework is applied to the study of gauge invariant cosmic perturbations and phenomenological consequences are worked out.   \\

{\bf 3) LQC extension of the inflationary scenario.} The inflationary scenario occupies the leading position in accounting for the origin of the cosmic inhomogeneities observed in the Cosmic Microwave Background (CMB) and  large scale structure. This success is mainly rooted in the economy of assumptions, the elegant mechanism that originates the  {\em cosmic inhomogeneities from  vacuum quantum fluctuations}, a subtle interplay between quantum mechanics and classical gravitation, and particularly  the non-trivial agreement with observations. Inflation is however an effective theory, and it is expected that a more fundamental theory will complete it. Examples of open questions that the more complete theory should answer are: What is the nature of the scalar inflaton field? Is there a single or several fields, like in multi-field models? What is the specific shape of the inflaton potential? These questions originate in particle physics, and unfortunately at these stages LQC does not have much to contribute. There are, in addition, important issues related to gravitation: What is the evolution of the space-time before inflation? In General Relativity the big bang singularity is unavoidable in inflationary scenarios \cite{bgv}. Is there a quantum gravity scenario in which the singularity is resolved {\em and} in which the evolution finds an inflationary phase compatible with observations generically, i.e. {\em without a fine-tuning of its parameters}? Such a scenario would allow to extend the inflationary space-times all the way back to the Planck era. Moreover, one could then use the quantum theory of cosmological perturbation on quantum space-times described in section~\ref{sec:3}, to extend the analysis of cosmic inhomogeneities to include Planck scale physics.

Section~\ref{sec:4} will review the arguments showing that such an extension is possible in LQC, where one can construct a {\em conceptual} completion of the inflationary theory from the quantum gravity point of view, in which Planck scale physics can be included in the study of cosmological perturbations. The importance of this extension goes, however, beyond the conceptual domain and may open a  window for phenomenological consequences.
\\

To summarize, this chapter  will review recent advances in the completion of the quantization program underlying LQG when restricted to the cosmological sector. We shall explore how the singularity of the homogeneous background is avoided, and how the abstract theoretical framework can descend down to make contact with phenomenology. Although many open issues still remain, at the present time there is a solid body of knowledge, based on a rigorous mathematical framework. These combine with analytical and numerical techniques, and provide an avenue from the big bang singularity resolution to concrete observation of the CMB and galaxy distributions. 

Due to space restrictions, there are some topics that we shall not cover in this chapter, such as the path integral formulation and its relation with spin foams \cite{ach}, spin foam cosmology \cite{vidotto}, the Gowdy models \cite{hybrid1,hybrid2,hybrid3,hybrid4,hybrid5}, nor numerical issues \cite{brizuela}. We do not provide either a review of all the existing ideas to study LQC effects on cosmic perturbations. See \cite{pert_tensor1, ns_inflation, barrau1, barrau2, barrau3, barrau4, bojowald&calcagni, barrau5, madrid, wilson-ewin} for different approaches to that problem.  Further information can be found in the reviews \cite{asrev}, \cite{lqcreview}, \cite{singh-numerical} and \cite{calcagni}.

Our convention for the metric signature is $-+++$, we set $c=1$ but keep $G$ and $\hbar$ explicit in our expressions, to emphasize gravitational and quantum effects. When numerical values are shown, we use Planck units.

\section{Quantization of cosmological backgrounds}
\label{sec:2}
In this section we shall consider the quantum theory of the homogeneous background within the context of Loop Quantum Cosmology. First we shall discuss what it means for a cosmological model to be quantized, or to use the standard nomenclature, to define a {\em quantum cosmology}. Just as with the quantization of any mechanical system such as the hydrogen atom, the first step is to cast the model to be quantized in a Hamiltonian language. That is, one has to identify configuration variables $q^i$ and their corresponding momenta $p_j$, with the property that the Poisson bracket is $\{q^i,p_j\}=\delta^i_j$. 
The next step in the quantization process is to find a Hilbert space ${\cal H}$ and
operators $\hat{q}^i$ and $\hat{p}_j$ satisfying
$[ \hat{q}^i, \hat{p}_j ]=i\hbar\, \delta^i_j$. Then one has to define an operator
$\hat{H}$ corresponding to the Hamiltonian (and to other physically relevant observables),
in order to define dynamics through the Schr\"odinger equation: 
$-i\hbar\,\partial_t \Psi= \hat{H}\Psi$.

In the case where the classical system under consideration is a {\em totally constrained system}, instead of a Hamiltonian $H$ defining dynamics, both the classical description and the corresponding quantization are more subtle.
Here the dynamical variables are subject to a constraint ${\cal C}(q,p)=0$. Furthermore, there is no Hamiltonian defining dynamics, and the canonical transformations generated by the constraint ${\cal C}$ are interpreted as {\em gauge}. That is, points on the phase space connected by a canonical transformations generated by the constraint are physically equivalent. Thus, the curve on phase space made out of all the physically equivalent points represents a {\em gauge orbit} and can be identified with a point on the true, {\em physical} phase space. Observables will be those functions $f(q,p)$ that are constant along the gauge orbits (i.e. satisfying $\{f,{\cal C}\}=0$).
Since there is no true dynamics, the system is said to posses a {\em frozen dynamics}. A natural question is whether one can extract some `dynamics' from the frozen formalism. In some cases, one can use one of the variables (or an appropriately selected function) as an internal time $T(q,p)$, with respect to which the gauge orbit can be described in terms of a relational dynamics (that is, where the `dynamics' is described by correlations between the variable $T$ and the rest of the variables). 

Let us now review the quantization process when we have a totally constrained system. The first step is to define a {\em kinematical Hilbert space} ${\cal H}_{\mathrm{kin}}$. This space serves as an arena for the implementation of the constraint, that is now required to
be represented as a self-adjoint operator $\hat{\cal C}$ on  ${\cal H}_{\mathrm{kin}}$. Not all states in the kinematical Hilbert space are regarded as physical. The condition that selects those physical states was put forward by Dirac and has the form,
\be
\hat{\cal C}\cdot\Psi_{\mathrm{phy}}=0\, .\label{dirac-cond}
\ee 
Once one has found the physical states $\Psi_{\mathrm{phy}}$ (that might belong to 
 ${\cal H}_{\mathrm{kin}}$ or not), one needs to specify an inner product 
 $\langle\cdot|\cdot\rangle_{\mathrm{phy}}$ in order to construct ${\cal H}_{\mathrm{phy}}$,
the {\em physical} Hilbert space. Physical observables will be operators $\hat{F}$ that leave the space of physical states invariant. This translates into the condition 
$[\hat{F},\hat{\cal C}]=0$. In some cases, when there is an internal time variable $T$, one can recast the Dirac condition (\ref{dirac-cond}) as an `evolution' equation where $T$
plays the role of time, as in the Schr\"odinger equation.

One interesting feature of the simplest cosmological models is that they are totally constrained systems, so the general framework we have outlined is applicable.  
Even more, one can complete the quantization program and obtain a complete physical description where a massless scalar field $\phi$ plays the role of internal relational time. One can then pose physical questions pertaining to observables of cosmological interest, such as the Hubble parameter and curvature scalars. Interestingly,
for the simplest models, one can indeed find {\em two} different, inequivalent, quantizations. The first one corresponds to the so-called Wheeler-De Witt (WDW) quantization that was put forward by De Witt and Misner in the 60's. The second quantization corresponds
precisely to the one we shall here consider in detail, known as loop quantum cosmology.
As we shall describe in more detail later, the basic difference between these two programs corresponds to the choice of kinematical Hilbert space ${\cal H}_{\mathrm{kin}}$. The choice made by De Witt and others was, in a sense, the most natural one, resembling the Schr\"odinger quantum mechanics that has been so useful to describe many physical systems. On the other hand, the choice one makes in LQC is somewhat exotic from the perspective of standard quantum mechanics, but is selected when the underlying symmetries pertinent to the gravitational field are seriously taken into account.

The second and physically most important difference between these two representations 
is that their predictions regarding the fate of the classical singularity are radically different. While the WDW theory predicts that the singularity remains, as defined by the behavior of the expectation values of physically relevant operators such as energy density, in the case of LQC the singularity is  generically avoided. Instead of a big bang (or big crunch) one has a bounce connecting a contracting branch with an expanding one; the energy density and curvature scalars are bounded from above, so that physics is well defined throughout the intrinsic dynamical evolution of the quantum state describing the universe.

Let us now briefly describe the structure of the remainder of this section.
In the first part, we study in detail the $k$=0 FLRW model with vanishing cosmological constant, and discuss some of its main features. In the second part we discuss other models. The first one we consider is the closed $k$=1 model also without a cosmological constant. Next, we briefly discuss $k$=0 FLRW models with a cosmological constant and some anisotropic models. In the third part, we introduce the so called effective equations. We give a brief introduction to the subject and discuss in detail the case of $k$=0 FLRW model. Next we consider the $k$=1 case, followed by a discussion of anisotropic effective space-times, including the Bianchi I, II and IX models.

\subsection{$k=0$ FLRW, singularity resolution}
\label{sec:2.a}

The simplest model that one can consider is a $k$=0 homogeneous and isotropic 
FLRW cosmological model  foliated by 3-manifolds $\Sigma$ that are topologically ${\mathbb{R}}^3$.
In order to find a Hamiltonian description for the model, we have to start with
an action principle. Due to
homogeneity, the action is not well defined unless one introduces and fixes a
fiducial cell ${\cal V}$. This will play the role of a co-moving volume. 
We can introduce a flat fiducial metric
$\q_{ab}$ on $\mathbb{R}^3$ with respect to which the coordinate
volume of ${\cal V}$ is $\V = \int_{\cal V}\, \sqrt{\q}\, \d^3\! x$. 
Without loss of generality, in what follows we shall set $\V=1$. 
The flat FLRW spacetime is described by the metric
\be
\d s^2 = - N^2 \d t^2 + a(t)^2 \d {\bf x}^2
\ee
where $N$ is the lapse function, $\q \leftrightarrow \d {\bf x}^2$ is the flat fiducial
metric, and $a$ is the {\em  scale factor} of the universe. 
Now, the action principle is 
%
\be \label{action} S
= \nonumber \f{1}{16 \pi G} \, \int \d t\,\int_{\cal V} \d^3\! x \sqrt{|g|}\, R =
\nonumber \f{1}{16 \pi G} \, \int \d t\, N\, a^3\,  R ~. 
\ee
with $R$ the scalar curvature of the spacetime.
The gravitational part of the 
phase space consists of $a$ and its conjugate momenta that is found to be:
$$P_a = - \f{3}{4 \pi G N} \, a \,\dot a\, .$$
In this simplest model, the matter we shall consider is a homogeneous 
massless scalar field $\phi$.
The action for such a field is: 
$$ S_{\rm matt}=\f{1}{2}\int\d t\,\f{a^3\dot{\phi}^2}{N}\, .$$
From this, the momenta $p_{(\phi)}$ associated to the scalar field is 
$p_{(\phi)}= \f{\dot{\phi}\,a^3}{N}$, and the Hamiltonian constraint that defines the `dynamics' is then,
\be
\label{WDW-HC}
{\cal C}_{\rm tot}= \f{2\pi G}{3}\f{P_a^2}{a} - \frac{1}{2}\f{p_{(\phi)}^2}{a^3}\approx 0\, .
\ee
To summarize, the phase space is four dimensional with coordinates $(a,P_a;\phi,p_{(\phi)})$,
satisfying $\{a,P_a\}=1$ and $\{\phi,p_{(\phi)}\}=1$. In the standard Wheeler-De Witt approach, the next step is to consider the kinematical Hilbert space to consist of
`wavefunctions' $\Psi_{\mathrm{wdw}}=\Psi(a,\phi)$ of the `configuration' variables $(a,\phi)$. In this case, the operators are represented in the usual fashion, as:
$\hat{a}\cdot\Psi(a,\phi)= a\Psi(a,\phi)$ and $\hat{P}_a=-i\hbar\partial_a\Psi(a,\phi)$, and similarly for the other variables.
Then, one promotes the constraint (\ref{WDW-HC}) to an operator, and finds
solutions to the Dirac condition (\ref{dirac-cond}). This has been described in detail
in \cite{aps3,acs}.

In order to define the corresponding phase space in loop quantum cosmology, we need to follow some more steps. The first one is that one needs to introduce a new set of variables for the gravitational degrees of freedom. As explained in Sahlmann's contribution to this volume, loop quantum gravity, and consequently LQC is based in a connection $A$ and its corresponding momenta $E$, a generalization of the magnetic potential and electric field of electromagnetism.
Let us then write the phase space in terms of these so-called  Ashtekar-Barbero
variables. First, introduce a fiducial triad $\e^a_i$ and co-triad
$\ow^i_a$ compatible with $\q_{ab}$. The conjugate phase space variables are the
$SU(2)$ connection $A^i_a = \Gamma^i_a + \gamma K^i_a$ and the
densitized triad $E^a_i$ satisfying
\be
\{ A^i_a(x),E^b_j(y)\} = 8 \pi G\, \gamma\, \delta^b_a \delta^i_j \delta^3(x,y)\, .
\ee
Here $\Gamma^i_a$ is the spin connection measuring the intrinsic
curvature (which vanishes in the $k=0$ model), $\gamma$ is the
Barbero-Immirzi parameter and $K^i_a$ is the extrinsic curvature
1-form related to the extrinsic curvature $K_{ab}$ as $K^i_a = e^{b i} K_{ab}$, with $e^a_i$  the un-densitized triad.
Due to the underlying symmetries of the homogeneous isotropic spacetimes we are considering, these variables can be written as \cite{abl}
\be\label{AE_defs}
A^i_a \, = \, c  \, \ow^i_a \quad ; \quad
E^a_i \, = \, p \, \sqrt{\q} \, \e^a_i \, .
\ee
Thus, the dynamical variables in the isotropic cosmological regime are $p$ and $c$.
The relationship between the `triad' $p$ and
the scale factor is, 
\be
\label{pa2} |p| =  a^2 \, . 
\ee
%
The connection component gets related to the rate of change of scale
factor as 
\be
\label{cdota} c = \gamma \,  \f{\dot a}{N}\, , 
\ee
holding only for the physical solutions of General Relativity (GR).
The gravitational part of the phase space is  characterized by the conjugate  variables $c$
and $p$ satisfying:
\be
\{c,p\} = \f{8 \pi G \gamma}{3}\, .
\ee
and the complete phase space has coordinates $(c,p;\phi,p_{(\phi)})$. The dynamics thus fund in
the Hamiltonian language is completely equivalent to the standard description based in Einstein's equations. To see that, one can find the Hubble parameter $H=\dot{a}/{a}=\dot{p}/(2p)$ by computing $\dot{p}=\{p,{\cal C}\}$, where ${\cal C}$ is now written in terms of the variables $(c,p;\phi,p_{(\phi)})$ (see (\ref{grav-const-cp}) below). 
From there one can write the standard Friedman equation: $H^2=\f{8\pi G}{3}\, \rho$, with $\rho=p_{(\phi)}^2/2V^2$.

Let us now consider the issue of quantization. As previously discussed, the choice of kinematical Hilbert space in LQC is different from the WDW case. That is, we do not expect to represent $\hat{c}$ and $\hat{p}$ as multiplication and derivation, for example.
The idea instead is to construct a quantum theory that
is closest to the quantization used in loop quantum gravity, as discussed for instance in \cite{lqg}. This means in particular a different choice of kinematical Hilbert space.
Recently this {\it polymeric} quantization for cosmological models
has been shown to be unique when invariance under diffeomorphisms is imposed \cite{ach4} (in complete analogy with the corresponding results in full LQG \cite{lost,cf}).
The new strategy is the following. Instead of re-writing the Hamiltonian constraint
(\ref{WDW-HC}) in terms of the $(c,p)$ variables, one starts with the full expression of the Hamiltonian constraint, in terms of variables $A$ and $E$. Then, one uses the simplification given by Eq.(\ref{AE_defs}). 
As it turns out, the choice of the polymeric Hilbert space as the kinematical arena for the implementation of the constraint --following the LQG route to quantization-- has the important feature that it does {\em not} admit the $\hat{c}$ operator. That is, only exponential functions of the gravitational connection $c$ such as
\be
h_k^{(\lambda_c)} = \cos (\lambda_c \, c/2) \mathbb{I} + 2
\, \sin  (\lambda_c \, c/2) \tau_k
\ee
become well defined. These objects have the geometrical interpretation of being the  `holonomies', or parallel transports of the connection $A$.
These functions generate an algebra of so-called almost periodic functions whose
elements are of the form $\exp(i \lambda_c \, c/2)$. 

The resulting
kinematical Hilbert space is then $L^2({\mathbb R}_{\mathrm{Bohr}},\d
\mu_{\mathrm{Bohr}})$, a space of square integrable
functions on the Bohr compactification of the real line. Beside the name of the space,
it is straightforward to understand the nature of this space.
For instance, the eigenstates of $\hat p$, labelled by $|\mu\rangle$,
satisfy $\langle \mu_1|\mu_2 \rangle = \delta_{\mu_1,\mu_2}$. 
This is to be contrasted with the usual Schr\"odinger representation where, instead of the
Kronecker delta one has the Dirac delta.

In particular, this eigenstates are
{\it normalized} and constitute a basis for the kinematical Hilbert
space ${\cal H}_{\mathrm{poly}}$. This constitutes the main difference from the standard Schr\"odinger representation
where the eigenstates of momentum $\hat{p}\,|\mu\ra = \mu\,|\mu\ra$ are {\it not} normalized and satisfy $\la\nu|\mu\ra = \delta(\mu,\nu)$. Note also that this plane waves states are {\it not} a basis for the $L^2(\mathbb{R},\d x)$ Hilbert space. 

There exists an important result in mathematical physics stating that for a finite dimensional phase space, the Schr\"odinger Hilbert space is the only choice of representation of the
canonical commutation relations, satisfying some regularity conditions. This result goes under the name of the Stone-Von Neumann uniqueness theorem \cite{afw}. Thus, one could have
imagined that, since the system has a finite number of degrees of freedom, both the WDW and the LQC representations should be equivalent. However, that expectation is not realized.
The {\it polymeric} representation used in LQC and the standard one are unitarily inequivalent. This is due to a crucial property of the LQC operators, implying that the polymer quantum mechanics does not posses some of the regularity conditions that go into the hypothesis of the Stone-Von Neumann theorem. 
To explore those propertie further, let us consider the
action of the two fundamental operators on the eigenstates $|\mu\rangle$,
\be
\label{p_act} \hat p\,| \mu \rangle = \f{8 \pi \gamma \lp^2}{6}
\mu\, |\mu \rangle \quad ; \quad 
{\widehat{\exp(i \lambda_c \, c/2)}}\,
|\mu\rangle = |\mu + \lambda_c \rangle
\ee
Note that the `displacement' operator ${\widehat{\exp(i \lambda_c \, c/2)}}$ is not continuous when $\lambda_c\to 0$, since the states $|\mu\rangle$ and $|\mu+\lambda_c\rangle$ are always orthogonal to each other, for all values of $\lambda_c > 0$. Also note that a basis of the polymer Hilbert space is uncountable as the label $\mu$ for the eigenstates can take any value in the real line.
%
%

In order to obtain the quantum constraint the key step is to rewrite the classical
gravitational constraint with field strength $F_{ab}^i$ as,
\be \label{eq:cgrav}
C_{\mathrm{grav}} = - \gamma^{-2} \int_{\cal
V} \d^3 x\,  \epsilon_{ijk} \,\f{E^{ai}E^{bj}}{\sqrt{|\det E|}}\,
F_{ab}^i ~ 
\ee 
Further, one writes the field strength $F_{ab}^i$ in terms of holonomies and triads and then
quantize (where we have chosen $N=1$). The matter part of the constraint is quantized in the
regular way, where the Schr\"odinger representation is used. A further simplification is to choose $N=a^3=V$ from the very beginning. If we
rewrite the line element with this choice we have $\d s^2=-a^6\d\tau +  a^2 \d {\bf x}^2$, for which
the classical constraint reads,
\be
 p_{(\phi)}^2 - \f{3}{4\pi G\gamma^2}\, p^2 c^2 =0\, . \label{grav-const-cp}
\ee
With this choice, the gravitational constraint has then the form,
\be
\label{eq:cgrav2}
{\cal C}_{\mathrm{grav}} = - \gamma^{-2}   {\epsilon^{ij}}_{k} \,\e^{a}_{i}\e^{b}_{j}\,
F_{ab}^k ~ 
\ee 
The field strength can be classically written in terms of a trace of
holonomies over a square loop $\Box_{ij}$, considered over a face
of the elementary cell, with its area shrinking to zero: 
\be
\label{F} F_{ab}^k\, = \, -2\,\lim_{Ar_\Box
 \rightarrow 0} \,\, \Tr\,
\left(\f{h^{(\lambda_c)}_{\Box_{ij}}-1 }{\lambda_c^2} \right)
\,\, \tau^k\, \ow^i_a\,\, \ow^j_b\, = \,\lim_{\lambda_c
 \rightarrow 0}   {\epsilon^k}_{ij}\, \ow^i_a\,\, \ow^j_b\,\left(
 \f{\sin^2{\lambda_c c}}{\lambda_c^2}\right) 
\ee
with
\be
h^{(\lambda_c)}_{\Box_{ij}}=h_i^{(\lambda_c)} h_j^{(\lambda_c)}
(h_i^{(\lambda_c)})^{-1} (h_j^{(\lambda_c)})^{-1}\, . \ee
%
%
Since the underlying geometry in the quantum theory resulting
from LQG is discrete, the loop $\Box_{ij}$ can be
shrunk at most to the area which is given by the minimum eigenvalue of the area operator in LQG:
$\Delta = \tilde\kappa\, \lp^2$ with $\tilde\kappa$ of order one. Note that it has been standard in the LQC literature to choose $\tilde\kappa= 2 \sqrt{3} \pi \gamma$  \cite{abl}, but
it can also be taken as a parameter to be determined \cite{acs}.
The area of the loop with respect to the physical metric is $\lambda_c^2 |p|$. Requiring
the classical area of the loop $\Box_{ij}$ to have the quantum area gap
as given by LQG, we are led to set
$\lambda_c = \sqrt{\Delta/|p|}$. Since $\lambda_c$ is now a
function of triad, the action of $\exp(i \lambda_c(p) c)$ becomes
complicated on the states in triad ($\mu$) basis. However, its
action in volume ($\nu$) basis is very simple: it drags the state
by a unit affine parameter.

It is then convenient to introduce the variable
$\b := \f{c}{|p|^{1/2}}$,
such that $\lambda_c c = \lambda_\b \b$ where $\lambda_\b := \sqrt{\Delta}$ is the
new affine parameter. Note that $\b$ is  conjugate variable to $\nu$,
satisfying  $\hbar \{\b,\nu\} = 2$, where $\nu$ labels the eigenstates of
the volume operator
\be
\hat V \, |\nu\rangle = 2 \pi \lp^2 \gamma |\nu| \, |\nu\rangle ~.
\ee
The action of the exponential operator then becomes very simple:
\be
\widehat{\exp(i \lambda_c c/2)} \, |\nu\rangle ~ = ~\widehat{\exp(i \lambda_\b \b/2)}
\, |\nu\rangle ~ = ~ |\nu +  \lambda_\b\rangle ~.
\ee
In what follows we shall only consider $\lambda_\b$ a constant and shall only denote it by
$\lambda$.
Further, all of the identities used to write classical constraint
in terms of holonomies remain unaffected and the 
quantum constraint operator on wave functions $\tilde{\Psi}(\nu,\phi)$ of $\nu$ and $\phi$ is obtained
\be \label{hc4}
\partial_\phi^2\, \tilde{\Psi}(\nu,\phi) = 3\pi G\, |\nu|\,
\f{\sin\lambda\b}{\lambda}\, |\nu|\, \f{\sin\lambda\b}{\lambda}\,
\tilde{\Psi}(\nu,\phi)\, 
\ee

Writing out the explicit action of operators
$\sin \lambda\b$, (\ref{hc4}) simplifies to:
\ba \label{hc5} \partial_\phi^2 \,\tilde{\Psi} (\nu, \phi) &=& 3\pi
G\, \nu\, \f{\sin\lambda\b}{\lambda}\, \nu\,
\f{\sin\lambda\b}{\lambda}\, \tilde{\Psi}(\nu,\phi) \nonumber\\
&=&\f{3\pi G}{4\lambda^2}\, \nu \left[\, (\nu+2\lambda)
\tilde\Psi(\nu+4\lambda) - 2\nu \tilde\Psi(\nu) + (\nu -2\lambda)
\tilde\Psi(\nu-4\lambda)\, \right]\nonumber\\
&=:& \Theta_{(\nu)}\, \tilde\Psi(\nu,\phi)\, . \label{Quant-Const}
\ea
The geometrical part, $\Theta_{(\nu)}$, of the constraint is a
difference operator in steps of $4\lambda$. 
\be C^+(\nu) \Psi(\nu + 4 \lambda) +
C^0 \Psi(\nu) + C^- \Psi(\nu - 4 \lambda) = \hat C_{\mathrm{matt}}
\, \Psi(\nu) \ee
where $C^{\pm}$ and $C^0$ are functions of $|\nu|$ \cite{aps2}. Note that the
equivalent of the Wheeler-De Witt equation is now a {\em difference} equation in
the geometrical variable, instead of a differential equation.

Then, physical states correspond to solutions to the quantum constraint (\ref{Quant-Const}), but they should also belong to the positive frequency part of the Hamiltonian constraint, and satisfy the `Schr\"odinger equation',
\be
-i\hbar \,\partial_\phi\, \Psi(\nu,\phi)=\sqrt{\Theta}\,\Psi(\nu,\phi)\, .\label{schr-eq}
\ee
Furthermore, they should be symmetric under $\nu\to -\nu$ and have finite norms under the inner product,
\be
\la\Psi_1|\Psi_2\ra = \sum_\nu\, \overline{\Psi}_1(\nu,\phi_o)\, |\nu|^{-1}\, \Psi_2(\nu,\phi_o)\, .
\ee
where the constant $\phi_o$ is arbitrary. As discussed above, these physical states can be interpreted as being solutions to `evolution equations' with respect to the internal time $\phi$. One can indeed define

The next step is to define relational observables that will have a clear interpretation in terms of $\phi$. For instance, one can define the operator $\hat{V}_{\phi_0}$, as the operator corresponding to {\em the volume $V$ when the scalar field takes the value $\phi_0$}. One can indeed define such Heisenberg operators by the standard prescription:
\be
\hat{V}|_{\phi_0}\cdot\Psi_{\mathrm{phy}}(\nu,\phi):=e^{i\sqrt{\Theta}(\phi-\phi_0)}\,\hat{V}\,e^{-i\sqrt{\Theta}(\phi-\phi_0)}\,\Psi_{\mathrm{phy}}(\nu,\phi) ,
\ee
where $\hat{V}$ is the standard Schr\"odinger operator (acting by multiplication in this case). In this manner one can define operators corresponding to matter energy density $\hat{\rho}_{\phi_0}$ and curvature scalars, all with a clear interpretation as being 
defined  at `time $\phi_0$'. 

As it turns out, one can perform a Fourier transform into the conjugate variable to $\nu$, and the resulting quantum constraint, a differential equation, can be solved exactly \cite{acs}. This allows one to have closed expressions for the expectation values of
the Heisenberg operators. Let us now describe this {\em solvable} model within LQC.

\vskip0.5cm

\noindent{\em Solvable loop quantum cosmology (SLQC)}. We now wish to work in
the $\b$ representation because the geometrical part of the quantum
constraint will also become a differential operator. Since
$\tilde\Psi(\nu,\phi)$ have support on the `lattice' $\nu = 4n\lambda$,
and since $\b$ is canonically conjugate to $\nu$, their Fourier
transforms $\Psi(\b,\phi)$ have support on the continuous interval
$(0, \pi/\lambda)$:
\be \Psi(\b,\phi) := \sum_{\nu=4n\lambda}\, e^{\ihalf \nu\b}\,\,
\tilde\Psi(\nu,\phi); \quad \hbox{\rm so that} \quad \tilde\Psi(\nu, \phi) =
\f{\lambda}{\pi}\, \int_0^{\pi/\lambda} \!\! \d\b\, e^{- \ihalf
\nu\b}\,\, \Psi(\b,\phi)\, .\ee
From the form (\ref{hc4}) of the constraint it is obvious that it
would be a second order differential operator in the
$\b$-representation. 
Let us set $\tilde\chi(\nu,\phi) = (\lambda/\pi
\nu)\tilde{\Psi}(\nu,\phi)$. Then, on $\chi(\b,\phi)$, the
constraint (\ref{hc4}) becomes
\be \label{hc7}
\partial^2_\phi \, {\chi}(\b,\phi) = \alpha^2
\, \left(\f{\sin
\lambda\b}{\lambda}\, \partial_\b\right)^2\,\, {\chi}(\b,\phi) \ee
with $\alpha=\sqrt{12\pi G}$.
Note however, that \emph{we did not} arrive at (\ref{hc7}) simply by
replacing $\b$ in the expression of the classical constraint by
$\sin\lambda\b/\lambda$ as is often done.
Rather, (\ref{hc7}) results directly from the
`improved' LQC constraint if one begins with a harmonic time
coordinate already in the classical theory.

To simplify the constraint further, let us set
\be\label{x}  x = \f{1}{\alpha}\, \ln (\tan
\f{\lambda\b}{2}),\quad \hbox{\rm or}\quad \b = \f{2}{\lambda}\,
\tan^{-1}\, (e^{\alpha\, x})\ee
so $x$ ranges  $(-\infty,\infty)$. Then (\ref{hc5}) becomes
just the Klein-Gordon equation
\be \label{hc8}\partial^2_\phi\,\, \chi(x,\phi) =
\partial_x^2\,\,\chi(x,\phi) =: -\Theta\,\, \chi(x,\phi)\, .\ee
The physical Hilbert space is given by positive frequency
solutions to (\ref{hc8}), i.e. satisfy
\be \label{hc9} -i \p_\phi \chi(x,\phi) = \sqrt{\Theta}\, \chi
(x,\phi)\, . \ee
We can again express the solutions in terms of their initial data
and decompose them into left and right moving modes $\chi(x,\phi)=
\chi_L(x_+)+ \chi_R(x_-)$. 
The physical states that we shall consider are positive frequency 
solutions of (\ref{hc8}). Since there are no fermions in the model, the
orientations of the triad are indistinguishable and $\chi(x,\phi)$
satisfy the symmetry requirement $\chi(-x,\phi) = -\chi(x,\phi)$.
Thus, we can write
$\chi(x,\phi) =  (F(x_+) - F(x_-))/\sqrt{2}$, where $F$ is an
arbitrary `positive frequency solution'. To be precise, $F(x)$
is a positive momentum function, i.e. with a Fourier transform that has support
on the positive axis. With such a choice, the solution to the constraint equation
become of positive frequency. The physical inner product is given by,
\ba
(\chi_1, \chi_2)_{\rm phy}
&=& -i\int_{\phi =\phi_0} [\bar\chi_1(x,\phi)\partial_\phi \chi_2(x,\phi)
-(\partial_\phi \bar\chi_1(x,\phi))\chi_2(x,\phi)] \, \d x\\
&=&i\int_{-\infty}^\infty [\partial_x \bar  F_1(x_+) F_2(x_+)
-\partial_x \bar  F_1(x_-) F_2(x_-)] \, \d x  ~. \label{inner-prod}
\ea
The action of the operator $\hat{p}_{(\phi)}$ on
physical states is then: $\hat{p}_{(\phi)}\, \chi = -i\hbar\,
\p_\phi \chi \equiv \sqrt{-\p_x^2}\; \chi$.
We can now compute the expectation values and fluctuations of fundamental
operator such as 
$\hat V|_{\phi{_o}}$, and $\hat p_{(\phi)}$. For {\it any} state on the physical Hilbert
space the expectation value of the volume operator at `time $\phi$' is given by
\be
\la\hat{V}\ra_\phi :=
(\chi, \hat{V}|_\phi\chi)_{\rm phy}
 = 2\pi \gamma \lp^2 (\chi ,|\hat \nu| \chi)_{\rm phy}
\ee
where $|\hat \nu|$ is the absolute value operator obtained from
\be
\hat \nu=-\f{2\lambda}{\alpha}\cosh(\alpha x)i\partial_x \, .
\ee
Using the inner product \eqref{inner-prod} the expectation value of $|\hat \nu|$ is given by
\ba
(\chi,|\hat \nu| \chi)_{\rm phy}
&=&i\int_{-\infty}^\infty [\partial_x  \bar F(x_+)( \hat \nu F(x_+))
-\partial_x \bar F(x_-)(-\hat \nu F(x_-))] \, \d x \nonumber \\
&=& \f{2\lambda}{\alpha}\int_{-\infty}^\infty 
[\partial_x \bar F(x_+)\cosh(\alpha x)\partial_x F(x_+)
+\partial_x\bar F(x_-)\cosh(\alpha x)\partial_x F(x_-)]\, \d x \nonumber \\
&=& \f{4\lambda}{\alpha}\int_{-\infty}^\infty 
\left|\f{\d F}{\d x}\right|^2 \cosh(\alpha(x-\phi)) \, \d x \, .
\ea
From these
expressions one can find the expectation value of certain relational (Heisenberg) operators.
For instance, the expectation value for the volume operator, at time $\phi$, takes the form,
\be
\langle\hat{V}\rangle_\phi = V_+\,e^{\, \alpha \,\phi}
+V_-\,e^{-\alpha\,\phi} \label{v-exp} ~,
\ee
with, $V_\pm$ constants that depend on the details of the initial (normalized) wave-function:
\be
V_{\pm} = \frac{4 \pi \gamma \lp^2
\lambda}{\alpha}\int \left|\frac{\d F}{\d
x}\right|^2\,e^{\mp\alpha\, x} \d x
\ee
>From (\ref{v-exp}), it follows that the expectation value of the volume $\hat{V}|_\phi$ is large at both very early and late times and  has a
non-zero global minimum 
$$V_{\mathrm{min}} = 2 (V_+ V_-)^{1/2}\, .$$ 
The {\it bounce} occurs at time
$$
\phi_\b^V = (2\, \alpha)^{-1} \ln(V_-/V_+)\, .
$$  
Around $\phi = \phi_\b^V$, the expectation value of the volume $\langle
\hat V\rangle_\phi$ is symmetric. Thus we see that {\em all} states undergo
a {\em big bounce} that replaces the big bang (in which the volume goes to zero
as $\phi\to \pm\infty$). 
Note: In the case of the WDW quantization, the expected volume reaches zero as
$\phi\to \pm\infty$, so in this sense one still reaches the singularity.

Another important observable to consider is the energy density $\hat{\rho}|_{\phi_0}$. Interestingly, this quantity possesses  an absolute upper bound on the physical Hilbert space.
Let us now see how this bound for energy density  arises.
Fix any state $\chi (x,\phi) = (1/\sqrt{2}) (F(x_+) - F(x_-))$ in
$\Hp$. Let us work in the Schr\"odinger picture at a fixed instant
of time, say $\phi_0$. Then, it follows that $\rho=\la\hat{\rho}|_{\phi_0}\ra$ is
given by \cite{acs},
\be \label{rhobound} \rho = \f{3}{8\pi\gamma^2 G}\,\,
\f{1}{\lambda^2}\,\,\, \f{\left[\int_{-\infty}^{\infty}\! \d x |\p_x
F|^2\right]^2} {\left[\int_{-\infty}^{\infty}\! \d x |\p_x
F|^2\,\, \cosh (\alpha x) \right]^2}\, \ee
where the integrals are performed at $\phi=\phi_0$. Since $\cosh
(\alpha x) \ge 1$, it follows that the ratio of the the two integrals
is bounded above by 1. This immediately implies that there is an absolute bound
given by
\be
\la\hat{\rho}\ra_\phi \le \rcr \quad\quad {\rm with} \quad \quad
\rcr := \f{3}{8\pi\gamma^2 G}\, \f{1}{\lambda^2}\, .
\ee
It is interesting to note that this quantity depends inversely with the {\em loop quantum
geometry scale} $\lambda$. Thus, in the limit $\lambda\to 0$, where we expect to recover
the WDW theory, the density becomes unbounded. That is precisely what is found in the
complete quantization of the WDW theory \cite{acs}. 

Using the standard choice for $\lambda$ in LQC, namely $\lambda^2 =
4\pi\sqrt{3}\,\gamma\lp^2$, we obtain:\\
 $ \rcr = (\sqrt{3})/(32\pi^2\gamma^3 G^2 \hbar)\approx 0.41 \rho_{\rm Pl}$, where we have used the standard choice for $\gamma$ coming from the black hole entropy computation in LQG where $\gamma=0.237$.

In a similar fashion, it is straightforward to see that one can also bound the corresponding operator for the `Hubble parameter operator' $\hat{H}|_{\phi}$ for physical states. In this case the bound takes the form,
$\la \hat{H}\ra_{\phi}< 1/(2\lambda\gamma)$. Note that, just as in the case of energy density, the bound on the Hubble parameter is inversely proportional to the loop quantum cosmology scale $\lambda$. In the limit $\lambda \to 0$, the corresponding quantity becomes unbounded.

Let us now summarize the main features of the complete quantization of this simple cosmological model
\begin{enumerate}

\item The bounce is not restricted to semi-classical states but occurs for
states in a dense sub-space of the physical Hilbert space. 

\item There exists a supremum of the expectation value for the
energy density. This maximum allowed density is $\rcr = \sqrt{3}/(32
\pi^2 \gamma^3 G^2 \hbar)$. 
We note that existence of an
absolute maximum of the energy density in this cosmological model
implies a non-singular evolution, in terms of physical quantities.
The singularity is therefore, resolved.

\item When curvatures become much smaller than the Planck curvature
(or for $\rho \ll \rcr$)
the expectation values of the Dirac observables agree with the
values obtained from classical GR.

\item For states which are semi-classical at late times, i.e. those
which lead to a large classical universe, the backward
evolution leads to a
quantum bounce in which the energy density of the field becomes arbitrarily close to
$\rcr \approx 0.41 \rho_{\mathrm{Pl}}$.

\item States that evolve to be semiclassical at late times, as determined by the
dispersion in canonically conjugate observables, have to evolve from
states that also had semiclassical properties  before the bounce (even when there might be  
asymmetry in their relative fluctuations without affecting semiclassicality)
\cite{cs2,kp2,cm1}. Semiclassicality is preserved to an amazing degree across the
bounce.

\end{enumerate}

This concludes our discussion of the quantization of the homogeneous background in the case that the matter content is a massless scalar field. This is the simplest isotropic model and is completely solvable. The question now is how to generalize these results for other
isotropic and anisotropic models. That will be subject of the next subsection.

\subsection{Other cosmologies}
\label{sec:2.b}

\subsubsection{$k$=1 FLRW}
\label{sec:2.b.1}
 
There are several generalization one might consider away from the $k$=0, $\Lambda$=0, FLRW cosmology. The simplest case is to consider the $k$=1 FLRW cosmological model \cite{apsv,warsaw1,ck2}. Even when it is not phenomenologically favored, it is important since it represents a spatially closed model that 
in the classical theory has both an expanding and a contracting phase continuously joined by a `recollapse' point where $H=\f{\dot{a}}{a}=0$. Therefore, it is an important test if one can recover
the classical recollapse from the quantum theory. 

The spacetimes under consideration are of the form $M=\Sigma\times \mathbb{R}$, where $\Sigma$ is a topological three-sphere $\mathbb{S}^3$. It is standard to endow $\Sigma$ with a fiducial basis of one-forms 
${}^o\!\omega^i_a$ and vectors ${}^o\!e^a_i$. The fiducial metric on $\Sigma$ is then ${}^o\!q_{ab}:=
{}^o\!\omega^i_a\,{}^o\!\omega^j_b\,k_{ij}$, with $k_{ij}$ the Killing-Cartan metric on su(2). Here, the
fiducial metric ${}^o\!q_{ab}$ is the metric of a three sphere of radius $a_0$. The volume of $\Sigma$ with respect to ${}^o\!q_{ab}$ will be denoted by $V_0=2\pi^2\,a_0^3$. We also define the quantity $\ell_0:=V_0^{1/3}$.
It can be written as $\ell_0=:\vartheta\, a_0$, where the  quantity $\vartheta:=(2\pi^2)^{1/3}$ will appear in many expressions.

The isotropic and homogeneous connections and triads can be written in terms of the fiducial quantities as follows,
\be
A_a^i=\f{c}{\ell_0}\,{}^o\!\omega^i_a\qquad ;\qquad E^a_i=\f{p}{\ell^2_0}\sqrt{{}^o\!q}\,{}^o\!e^a_i\, .
\ee
Here, $c$ is dimension-less and $p$ has dimensions of length. The metric and extrinsic curvature can be recovered from the pair $(c,p)$ as follows, $
q_{ab}=\f{|p|}{\ell^2_0}\,{}^o\!q_{ab}$, and $\gamma K_{ab}=\left(c-\f{\ell_0}{2}\right)\f{|p|}{\ell^2_0}\,{}^o\!q_{ab}$.
Note that the total volume $V$ of the hypersurface $\Sigma$ is given by $V=|p|^{3/2}$. 
>From here, one can calculate the curvature $F^k_{ab}$ of the connection $A_a^i$ on $\Sigma$ as,
$F^k_{ab}=\f{c^2-2\vartheta c}{\ell^2_0}\;{\epsilon_{ij}}^k\,{}^o\!\omega^i_a\,{}^o\!\omega^j_b$.
The only relevant constraint is the Hamiltonian constraint that has the form,
\be
{\cal C}_{\textrm{grav}}=-\f{3}{8\pi G\gamma^2}\,\sqrt{|p|}\left[(c-\vartheta)^2 + \gamma^2\vartheta^2\right]
\ee
It is convenient to also use the variables \cite{acs}: $\b:=c/|p|^{1/2}$ and $V=p^{3/2}$. The
quantity $V$ is just the volume of $\Sigma$ and $\b$ is its canonically conjugate,
$\{\b,V\} = 4\pi G\gamma$.
We can then compute the evolution equations of $V$ and $\b$ in order to find interesting geometrical scalars.
Then,
\be
\dot{V}=\{V,{\cal C}_{\textrm{grav}}\}= \f{3}{\gamma}\left(\b V - \vartheta V^{2/3}\right)
\ee
from which we can find the standard Friedman equation using the constraint equation ${\cal C}=
{\cal C}_{\textrm{grav}} + {\cal C}_{\textrm{matt}}\approx 0$ and ${\cal C}_{\textrm{matt}}=V\rho$, we have
$H^2:=\left(\f{\dot{V}}{3V}\right)^2=\frac{8\pi G}{3} \,\rho-\frac{\vartheta^2}{V^{2/3}}$.
 
The basic strategy of loop quantization, just as in the $k$=0 case,  is that the effects of quantum geometry are manifested by means of holonomies
around closed loops to carry information about field strength of the connection.  In order to define the quantum theory, taking again $N=a^3$, one can work in the $\nu$ representation and define operators associated to curvature and spin connection to arrive to a difference operator $\Theta_{(k=1)}$ of the form,
\ba
\partial^2_\phi \Psi(\nu,\phi) &=&
\Theta_{(k=1)}\, \Psi(\nu,\phi) \nonumber \\
&=& -\Theta
\Psi(\nu,\phi) + \f{3\pi G}{\lambda^2}\,\nu\left[\sin^2\left(\f{\lambda\vartheta}{\tilde{K}\nu^{1/3}}\right)+(1+\gamma^2)\left(\f{\lambda\vartheta}{\tilde{K}}\right)\right]\,\Psi(\nu,\phi)\, ,
\ea
with $\tilde{K}=2\pi\gamma\lp$. Numerical solutions of this equation were studied in detail in
\cite{apsv} for sharply peaked states, and were shown to posses not only a bounce very close to the
critical density $\rcr$, but also a recollapse at a density and volume very close to the classical
value. Thus, this model provides a very striking example of a quantum gravitational system that possesses satisfactory UV and IR behavior.
The relative dispersion of $\hat{V}|_\phi$ does increase but the increase is very small: For a
universe that undergoes a classical recollapse at $\approx$ 1 Mpc, a state that nearly saturates the
uncertainty bound initially, with uncertainties in $\hat{p}_\phi$ and  $\hat{V}|_\phi$ spread equally, the relative
dispersion in  $\hat{V}|_\phi$ is still $\approx 10^{-6}$ after some $10^{50}$ cycles \cite{apsv}.
The expectation values of volume has a quantum bounce which occurs at $\rho=\rcr$ up
to the correction terms of the order of $\lp^2/V^{2/3}_{\rm bounce}$. For universes that grow to macroscopic sizes, the correction is totally negligible. For example, for a universe which grows to a
maximum volume of $1Gpc^3$, the volume at the bounce is approximately $10^{117}\lp^3$.
On the other hand, the numerical simulations show that one indeed recovers the recollapse with
very large precision for semiclassical states that reach large volumes  \cite{apsv}. An important
lesson that this model teaches us is that energy density and curvature are the relevant quantities to define what the Planck scale is, and not the size of the universe at the bounce (that, as we have seen, can be very large in Planck units). One should also note that, while semiclassical states alternate between the Planck scale (UV) and the low density, large volume GR regime (IR), states that are `truly quantum' --or far from semiclassical--  might have a bounce at a density much lower than $\rcr$, and not grow to large volumes before recollapse.

There exists another quantization in which the curvature is not obtained by means of closed holonomies, but rather by approximating the {\it connection} by open holonomies, as is done in
anisotropic models with non-trivial curvature \cite{ck2}. The structure of the constraints is different but its quantum solutions have not been explored numerically yet. 

Let us comment on the quantization of the $k$=-1 case. Some early attempts to find such a quantization were put forward in \cite{k=-1,szulc}, but those efforts still suffer from some drawbacks, such as absence of essential self-adjointness. A quantization based in open holonomies as in \cite{ck2} is still to be constructed.

\subsubsection{FLRW with $\Lambda\neq 0$}
\label{sec:2.b.2}

The results found for a zero cosmological constant can be generalized to the case of a non-zero cosmological constant. For a mass-less scalar field and both signs of the constant, we have singularity resolution, in the sense that the big bag/crunch is replaced by a bounce, just as in the $\Lambda=0$ case. For simplicity we shall consider the $\Lambda<0$, $k$=0 case, but the results can be generalized to $k$=1 as well. The Hamiltonian constraint, for $N=1$, takes the form,
\be
{\cal C}=
\f{p_{(\phi)}^2}{2V}
-\f{3}{8\pi G\gamma^2}\,\b^2 V +\f{\Lambda}{16\pi G}\, V\approx 0\, .
\ee
One can solve the equations of motion and express the dynamics in terms of the scalar field $\phi$ as,
\be
V(\phi)=\f{\alpha\,p_{(\phi)}}{\sqrt{3|\Lambda|}}\;\f{1}{\cosh[\alpha(\phi-\phi_o)]}
\ee
With this, there is a big bang singularity in the past $\phi\to -\infty$ and a big crunch in the future, when $\phi\to \infty$. There is a point of recollapse, when the volume reaches its maximum
value $V_{\rm max}=({\alpha\,p_{(\phi)}})/({\sqrt{3|\Lambda|}})$, at $\phi=\phi_o$, with some resemblance to the $k$=1 case.The quantum constraint takes now the form,
\be
\partial_\phi^2\Psi(\nu,\phi) = -\Theta\,\Psi(\nu,\phi) - \f{\pi G\gamma^2|\Lambda|}{2}\,\nu^2 \Psi(\nu,\phi)\, ,
\ee
with $\Theta$ the operator corresponding to the $k$=0, $\Lambda$=0 case. The operator can be consistently defined, and numerically solved \cite{bp} to give a picture very similar to the $k$=1 case with vanishing cosmological constant. The big bang/crunch is replaced by a bounce, in such a way that a sharply peaked state goes through a series of bounces and recollapses in an almost periodic fashion.

Let us now consider the $\Lambda>0$ case. The solution to the classical equations is slightly different from the negative case and takes the form \cite{ap,kp1},
\be
V(\phi)=\f{\alpha\,p_{(\phi)}}{\sqrt{3|\Lambda|}}\;\f{1}{\sinh[\alpha(\phi-\phi_o)]}
\ee
This is qualitatively very different from the previous case. Now, an expanding solution with a big bang singularity at the past, $\phi\to-\infty$, reaches an infinite volume for a {\it finite} value of
$\phi$, namely when $\phi=\phi_o$. Similarly, there are contracting solutions that `start', for $\phi=\phi_o$, with an infinite volume and end in a big crunch singularity when $\phi\to \infty$.
At the point $\phi=\phi_o$, the proper time diverges and the matter density vanishes. One can see that one can actually continue the classical evolution past this `singular' point \cite{ap}. In the quantum theory, this new behavior manifests itself in the fact that the operator $\Theta_\Lambda$ fails to be essentially self-adjoint, and one has the freedom of choosing different self-adjoint extensions. Interestingly enough, for all of them, the evolution of semiclassical states is almost indistinguishable. Evolution is well defined past the point $\phi=\phi_o$ and the universe recollapses. As in all previous cases, the big bang/crunch singularity is replaced by a bounce. 

\subsubsection{Anisotropic Cosmologies.}
\label{sec:2.b.3.}

Isotropic loop quantum cosmology, as we have seen, enjoys a very robust formulation; one has complete mathematical control over the quantum theory, one can make physical predictions using analytical or numerical tools and can therefore draw conclusions about the behavior of a background isotropic quantum geometry. The same is not true for anisotropic solutions. While the quantum constraints have been formulated in several cases, one does not have full mathematical control regarding their time evolution, and one has not been able to solve, even numerically, their dynamical evolutions.
In this part we shall summarize the formulation of the quantum models as we currently understand them. 

Let us consider the spacetime of the form $M=\Sigma\times\mathbb R$ where $\Sigma$ is a spatial 3-manifold which can be identified by the symmetry group of the chosen model and is endowed with a fiducial metric ${}^oq_{ab}$ and associated fixed fiducial basis of 1-forms ${}^o\omega_a^i$ and vectors ${}^oe_i^a$. If $\Sigma$ is non-compact then  
we fix a fiducial cell, $\mathcal V$, adapted to the fiducial triads with finite fiducial volume. We also define $L_i$ which is the length of the $i$th side of the cell along ${}^oe_i$ and $\V=L_1L_2L_3$. We choose for compact $\Sigma$, $L_i=\V^{1/3}$ with $i=1,2,3$.

Since all of the models in which we are interested are homogeneous and, if we restrict ourselves to diagonal metrics, one can fix the gauge in such a way that 
$A_a^i$ has 3 independent components, $c^i$, and $E_i^a$ has 3 independent components, $p_i$,
\begin{equation}
A_a^i=\frac{c^i}{L_i}{}^o\omega_a^i\quad \textrm{and}\quad E_i^a=\frac{p_iL_i}{\V}\sqrt{{}^oq}{}\ ^oe_i^a
\end{equation}
where $p_i$, in terms of the scale factors $a_i$, are given by $|p_i|=L_iL_ja_ja_k$ ($i\neq j\neq k$).  Using $(c^i,p_i)$ for anisotropic models, 
the Poisson brackets can be expressed as $\{c^i,p_j\}=8\pi G\gamma\delta_j^i$. 
With this choice of variables and gauge fixing, the Gauss and diffeomorphism constraints are satisfied and the only constraint is the Hamiltonian constraint
\begin{equation}\label{FHC}
\mathcal C_H=\int_\mathcal V N\left[-\frac{\epsilon^{ij}_{\ k}E_i^aE_j^b}{16\pi G\gamma^2\sqrt{|q|}}\left(
F_{ab}^k-(1+\gamma^2)\Omega_{ab}^k\right)+\mathcal H_{\rm matter}\right]\textrm{d}^3x \, ,
\end{equation}
with $N$ the lapse function, $\mathcal H_{\rm matter}=\rho V$ and $\Omega_{ab}$ the curvature of the spin connection $\Gamma_a^i$ compatible with the triads.

Using a strategy similar to the isotropic case, the field strength $F_{ab}^k$ is given by
\begin{equation}
F_{ab}^k=2\lim_{Area_\square\rightarrow 0}\epsilon_{ij}^{\ \ k}\textrm{Tr}\bigg(\frac{h_{\square_{ij}}^{\mu^\prime}-\mathbb I}{\mu^\prime_i\mu^\prime_j}\tau^k\bigg){}^o\omega_a^i{}^o\omega_b^j \, .
\label{fs}
\end{equation} 
The strategy to choose the corresponding loops is slightly different from the isotropic case.
We take  $\mu_i^\prime=\bar\mu_i L_i$ where $\bar\mu_i$ is a dimensionless parameter and, by previous considerations,  is equal to $\bar\mu_i=\lambda\sqrt{|p_i|}/\sqrt{|p_jp_k|}$ ($i\neq j\neq k$) \cite{awe2,madrid-bianchi}.

For Bianchi II and IX models, this strategy fails because the resulting operator is not almost periodic. Therefore, we express the connection $A_a^i$ in terms of holonomies and then use the standard definition of curvature $F_{ab}^k$. 
The operators corresponding to the connection are given by \cite{awe3}
\begin{equation}
\hat c_i=\widehat{\frac{\sin\bar\mu_ic_i}{\bar\mu_i}}\, . \,\,
\end{equation}
Note that using this quantization method for flat FLRW \cite{aps3} and Bianchi I \cite{awe2} models, one has the same result as the direct quantization of curvature $F_{ab}^k$ (with proper identification of the
parameters), but for a closed FLRW it leads to a different quantum theory which is more compatible with the isotropic limit of Bianchi IX \cite{ck2, we, ck3}. We call the first method of quantization {\it curvature based quantization} and the second one {\it connection based quantization}.

In Bianchi II and Bianchi IX models the terms related to the curvatures, 
$F_{ab}^k$ and $\Omega_{ab}^k$, contain some negative powers of $p_i$ which are not well defined operators. To solve this problem we use the same idea as Thiemann's strategy,
\begin{equation}
|p_i|^{(\ell-1)/2}=-\frac{\sqrt{|p_i|}L_i}{4\pi G\gamma j(j+1)\tilde\mu_i\ell}\textrm{Tr}(\tau_i h_i^{(\tilde\mu_i)}\{h_i^{(\tilde\mu_i)-1},|p_i|^{\ell/2}\}) \, ,
\label{np}
\end{equation}
where $\tilde\mu_i$ is the length of a curve, $\ell \in (0,1)$ and $j\in \frac{1}{2}\mathbb{N}$ is for the representation.
Therefore, for these three different operators we have three different curve lengths ($\mu,\mu^\prime,\tilde\mu$) where $\mu$ and 
$\tilde\mu$ can be some arbitrary functions of $p_i$, so for simplicity 
we can choose all of them to be equal to $\mu^\prime$. On the other hand we have another free parameter in the definition of 
negative powers of $p_i$ where, for simplicity, we take $j=1/2$. Since the largest negative power of $p_i$  which appears in the constraint is $-1/4$, we will take $\ell=1/2$ and obtain it directly from Eq.(\ref{np}), and after that we express the other negative powers by them. 
The eigenvalues for the operator $\widehat{|p_i|^{-1/4}}$ are given by
\begin{equation}
J_i(V,p_1,p_2,p_3)=\frac{h(V)}{V_c}\prod_{j\neq i}p_j^{1/4}\, ,
\end{equation}
with
\begin{equation}
h(V)=\sqrt{V+V_c}-\sqrt{|V-V_c|},\,\, \textrm{ and } \,\,\, V_c=2\pi\gamma\lambda\ell_p^2\, .
\end{equation}

By using these results and choosing some factor ordering, we can construct the total constraint operator. Note that  different choices of factor 
ordering will yield different operators, but the main results will remain almost the same. By solving the constraint equation 
$\hat{\mathcal C}_H\cdot\Psi=0$, we can obtain the physical states and the physical Hilbert space $\mathcal H_{\rm phys}$. 
As a final step, one would need to identify the physical observables, that in our case would correspond to relational observables as functions of the internal time $\phi$. These steps have proven to be exceptionally difficult and have prevented from solving the resulting difference equations numerically, even for the simplest case of Bianchi I.
\\

\subsection{Effective Equations}
\label{sec:2.c}

When analyzing the numerical solutions of the $k$=0, $\Lambda$=0 FLRW model, the authors of \cite{aps3} noticed that sharply peaked states followed trajectories in the $(V,\phi)$ plane that have a bounce, and therefore do not satisfy the classical Einstein equations. Furthermore, they
realized that the expectation value of $\hat{V}|_\phi$ {\it does indeed} follow a trajectory 
that satisfies (to a very good approximation) some equations that are now referred to as the {\it effective equations}. As it turns out, these
effective equations can be derived from an effective Hamiltonian constraint ${\cal C}_{\rm eff}$.
The question that arises then is how to derive, from the quantum theory defined by a quantum constraint $\hat{\cal C}$, the effective Hamiltonian. A second question pertains to the domain
of validity of these effective equations. That is, for which states and in which regimes are these equations a good approximation to the exact quantum dynamics? As we shall see in this part, for the models that are well understood, effective equations describe very accurately
 the dynamics for appropriately defined semiclassical states.

In the case of models for which we do not posses the full quantum dynamics, one can expect that
the effective theory to describe very well the quantum theory for semiclassical states far from the `deep quantum regime' (where it is expected to fail). Thus, in the anisotropic Bianchi I, II and IX models, the effective description that we shall here consider provide a description in which the singularity is also replaced by a bounce.

Let us begin by briefly describing how one obtains this effective descriptions from the quantum theory. The idea is to employ the geometric formulation of quantum mechanics \cite{as}, which provides an appropriate formalism from which one can find the effective Hamiltonian constraint ${\cal C}_{\mathrm{eff}}$
by computing the expectation value $\la \hat{C}\ra_\psi$ of the quantum Hamiltonian constraint on an appropriately defined semiclassical state $\psi$. From that expression one can find the
effective equations of motion by replacing ${\cal C}_{\mathrm{eff}}$ in Hamilton's equations:
$\dot{q}=\{q,{\cal C}_{\mathrm{eff}}\}$ and $\dot{p}=\{p,{\cal C}_{\mathrm{eff}}\}$.

Let us now be more precise. 
In the geometric formulation of quantum mechanics the space of quantum states is seen as a symplectic space $\Gamma_Q$, equipped with
a symplectic structure $\Omega_Q$ that is given by the imaginary part of the Hermitian inner product
$\la\cdot,\cdot\ra$ on $\H$. For each observable $\hat{F}$ one can define a function $\bar{F}:=\la\hat{F}\ra$ on normalized states. There is a corresponding Hamiltonian vector field
for each function $X^\alpha_{\bar{F}}=\Omega_Q^{\alpha\beta}\partial_\beta\bar{F}$. There is an
interesting interplay between these vectors and the vector one would obtain by acting with the operator $\hat{F}$ on a state $\Psi$,
\be
(\hat{F}\Psi)^\alpha=i\hbar\,X^\alpha_{\bar{F}}|_\Psi
\ee
Furthermore, the commutator of observables in the Hilbert space and the corresponding {\it quantum} Poisson bracket $\{\bar{F},\bar{G}\}_Q:=\Omega^{\alpha\beta}_Q\partial_\alpha\bar{F}\partial_\beta
\bar{G}$ satisfy the relation,
\be
\la[\hat{F},\hat{G}]\ra = i\hbar\,\{\bar{F},\bar{G}\}_Q
\ee
Thus, quantum dynamics can just be seen as ordinary Hamiltonian dynamics on the quantum phase space $\Gamma_Q$, as defined by the corresponding vector field $X^\alpha_{\bar{H}}$. How can we relate then this quantum evolution with the classical evolution on the phase space $\Gamma$? The idea is to project the dynamics on $\Gamma_Q$ to $\Gamma$ by means of appropriate coordinate functions. To be precise, let us assume that the classical phase space $\Gamma$ has coordinates $(q^i,p_i)$. In the Hilbert space one has the corresponding operators $(\hat{q}^i,\hat{p}_j)$. Then, one can define the projection $\Pi:\Gamma_Q \to \Gamma$ as follows: $\Pi:\Psi \to (\bar{q}^i,\bar{p}_j)$. One can now, given a quantum dynamical trajectory $\Upsilon_t$ on $\Gamma_Q$, define the corresponding projected classical trajectory $\gamma_t$ in $\Gamma$ as: $\gamma_t=\Pi(\Upsilon_t)$. 
The question that arises then is whether one can find an {\it effective} Hamiltonian $H_{\rm eff}$, defined on the classical phase space (and therefore being a function of $(q^i,p_j)$ and possible some parameters), such that the trajectory 
$\gamma_t=(\bar{q}^i,\bar{p}_j)$ follows Hamilton's equations $\dot{\bar{q}}^i=\{q^i,H_{\rm eff}\}$ and $\dot{\bar{p}}_j=\{p_j,H_{\rm eff}\}$. For this conditions to be satisfied, one must choose a particular `initial state' in order to select a preferred trajectory $\Upsilon_t$. In practice one looks for something simpler. In the so called, `embedding approach', one seeks an embedding $\Gamma \to \bar{\Gamma}_Q\subset\Gamma_Q$ of the finite dimensional phase space into the infinite dimensional quantum space $\Gamma_Q$ that is
well suited to capture the quantum dynamics, in the sense that the dynamical evolution lies approximately within  $\bar{\Gamma}_Q$. To define $\bar{\Gamma}_Q$, for any given point $\gamma^o\in\Gamma$, where $\gamma^o=(q_i^o,p_i^o)$, one prescribes a quantum state $\Psi_{\gamma^o}$
 for all $\gamma^o\in\Gamma$. A first requirement is that the embedding should be such that 
 $q_i^o=\la\Psi_{\gamma^o}\,\hat{q}_i\,\Psi_{\gamma^o}\ra$ and $p_i^o=\la\Psi_{\gamma^o}\,\hat{p}_i\,\Psi_{\gamma^o}\ra$. The second condition is dynamical and
non-trivial; it requires that the quantum Hamiltonian vector field should be approximately tangent to
$\bar{\Gamma}_Q$. If this is satisfied, one can project the exact quantum evolution $\Upsilon_t$ to
$\bar{\Gamma}_Q$ to obtain  $\bar{\Upsilon}_t$, and from this, project down to $\gamma_t=\Pi(\bar{\Upsilon}_t)$. It is natural to regard, as a candidate for $H_{\rm eff}$, the expectation value of the quantum Hamiltonian on the embedded submanifold: 
$H_{\rm eff}(q_i^o,p_j^o):= \la\Psi_{\gamma^o}\hat{H}\Psi_{\gamma^o}\ra$. One should note that for the ordinary harmonic oscillator, coherent states represent an exact dynamical embedding. That is, the exact quantum evolution lies within $\bar{\Gamma}_Q$ and the effective Hamiltonian coincides with the classical one. There are no {\it quantum} corrections to the dynamics from these states.

Let us now consider some important cases in homogeneous loop quantum cosmology.

\subsubsection{$k$=0 FLRW cosmology}
\label{sec:2.c.1}

Using the geometric methods of quantum mechanics just described, one can write an
effective Hamiltonian which provides an excellent approximation to the
behaviour of expectation values of Dirac observables in the
numerical simulations \cite{vt}. The effective Hamiltonian will in principle also
have contributions from terms depending on the properties of the state such as its
spread. Effect of these terms
turns out to be negligible as displayed from the detailed numerical analysis \cite{aps2,apsv}. 
Thus, the effective Hamiltonian constraint is, for $N$=1,
\be \label{effham}
{\cal C}_{\rm eff}=\f{3}{8 \pi G\gamma^2} \, \f{\sin^2(\lambda\, \b)}{\lambda^2} V - 
{\cal C}_{\mathrm{matt}} \, ,
\ee
which leads to modified Friedman and Raychaudhuri equations on computing the
Hamilton's equations of motion (as we shall see below). Using (\ref{effham}) one can find that the
energy density $\rho = H_{\mathrm{matt}}/V$ equals
$3 \sin^2(\lambda\, \b)/(8 \pi G \gamma^2 \lambda^2)$. Since the latter reaches its
higher possible value when $\sin^2(\lambda\, \b)=1$, the density has a maximum given by
\be
\rho_{\rm max}= \f{3}{8 \pi G \gamma^2 \lambda^2}\, ,
\ee
Thus, we see that that the maximum energy density obtained from the effective Hamiltonian is
identical to the supremum $\rho_{\mathrm{sup}}$ for the density operator in $k$=0, LQC. The difference is, of course, that in the effective dynamics every trajectory undergoes a bounce and reaches the maximum
possible density, while in the quantum theory not every state is close to the critical density at the quantum bounce. 

It is easy to solve for the dynamics defined by the effective Hamiltonian. The equations of motion
are found using the effective constraint: $\p_t F =: \dot{F} = \{F,{\cal C}\}$, with $t$ the cosmic time.
The only equation of motion different from the classical one (on the constraint surface) is
\be
\dot V=\frac{3}{\gamma\l}V\sin{(\l\b)}\cos{(\l\b)}\, ,
\ee
leading to the modified Friedman equation for the Hubble parameter
\be
H^2:=\left(\frac{\dot{a}}{a}\right)^2=\left(\frac{\dot{V}}{3V}\right)^2=\frac{8\pi G}{3}\,\rho\, \left(1-\frac{\rho}{\rcr}\right)\, ,\label{eff-fried}
\ee
where $\rho_{\rm max}=\frac{9}{2\alpha^2}\frac{1}{\l^2}$ is the scalar field density at the bounce.
For every trajectory there are quantum turning points at $\b =\pm\frac{\pi}{2\l}$, where $\dot{V}=0$,
corresponding to a bounce. Note that, at the bounce
$\ddot V\vert_{\beta =\frac{\pi}{2\l}}=2\,\alpha^2 V \rcr >0$,
so the bounce corresponds to a minimum of volume. Also, note that the Hubble parameter is absolutely
bounded $|H|\leq 1/(2\l\gamma)$, indicating that the congruence of cosmological observers can never
have caustics, independently of the matter content.

In the case of effective theories the proper time appears as a natural choice for an
evolution parameter, but one can always look for internal, relational notions of time.
Since, $\dot{\b}\le 0$ one can choose $\b$ as a relational time in the effective
theories, and consider the evolution with respect to $\b$. The advantage of this
election is that no external time variable is needed. Every trajectory,
that corresponds to $\b >0$, has a bounce
at $\b =\frac{\pi}{2\l}$, and this value tends to infinity as $\lambda\to 0$. 
 
In the effective theories, we consider the interval $\b\in [-\frac{\pi}{2\l},\frac{\pi}{2\l})$. One should note that all functions and observables in $\bar{\Gamma}_\l$ are periodic in $\b$ with period $\pi/\l$. It is then completely equivalent to regard the coordinate $\b$ as compactified on a circle. The solutions are defined for every $\t$ and are given by \cite{cv1},
\be
\cot{\l\b}=\frac{3\t}{\gamma\l}\, ,\ \ \
V_{\l}(\t )=\frac{\alpha}{3}\,p_{\phi}\,\sqrt{\gamma^2\l^2+9\t^2}\, ,
\ee
and
\be
\phi_{\l} (\t )=\phi_0 +\l\varphi +\frac{1}{\alpha}\ln{\frac{3\t+\sqrt{\gamma^2\l^2+9\t^2}}
{3\t_0+\sqrt{\gamma^2\l^2+9\t_0^2}}}\, ,
\ee
so that $\phi_{\l}(\t_0)=\phi_0 +\l\varphi$
and the initial condition approaches the classical one (for $\t=\t_0$) as $\l\to 0$.
Note that $\phi_{\l}(0)\to\frac{{\rm sgn}\b}{\kappa}\ln{\l}$ as $\l\to 0$.
Let us now consider an intrinsic description of the dynamics in terms of the scalar field.
One can solve $V$ as a function of $\phi$, 
\be
V_{\l}(\phi )=V_+e^{\alpha ({\rm sgn}\b )  (\phi -\phi (\t_0))}
+V_-e^{-\alpha ({\rm sgn}\b )(\phi -\phi (\t_0))}\, ,
\ee
where $V_+=\frac{1}{2}(V_0+\sqrt{V_0^2-\beta^2})$ and $V_-=\frac{\beta^2}{4}(V_+)^{-1}$,
where $V_0=V(\phi (\t_0))$, and  $\beta =\frac{1}{3}\gamma\l\alpha p_\phi$.
Note that the effective theory recovers the quantum dynamics of $\la\hat{V}\ra|_\phi$ {\it exactly},
for all states of the physical Hilbert space. That is, there are no further quantum corrections to
the dynamics of $V_\l(\phi)$.

With this, one can see that the effective theory defines an effective 
homogeneous and isotropic spacetime metric, that takes the form,
\be
(\d s^2)_{\rm eff} = - \d\t + a^2(\t)_{\rm eff}\; \d {\bf x}^2
\ee
with $a(\t)_{\rm eff}= \left(\frac{\alpha}{3}\right)^{\f{1}{3}}\,\f{p_{\phi}}{\V}\,
(\gamma^2\l^2+9\,\t^2)^{\f{1}{6}}$.
It is trivial to see that in the $\lambda\to 0$ limit, one recovers the classical spacetime
metric satisfying Einstein equations.

As we have seen, the quantum corrections captured by the effective Hamiltonian modify the 
Friedman equation in a non-trivial way, ensuring that quantum effects become important near
the Planck scale in such a way that a repulsive force is capable of stopping the collapsing universe and turn it around into an expanding phase. Let us explore a little bit more how this quantum repulsive force can be seen. First, a modified Raychaudhuri equation can be written \cite{ps},
\be
\f{\ddot{a}}{a}=-\f{4\pi G}{3}\,\rho\left(1-4\, \f{\rho}{\rcr}\right) -4\pi G\,P\left(
1-2\, \f{\rho}{\rcr}\right)\, .
\ee
It is also illustrative to write an equation for the rate of change of the Hubble parameter \cite{cs3},
\be
\dot{H}= -4\pi G (\rho + P)\left( 1-2\, \f{\rho}{\rcr} \right)\, .\label{dotH}
\ee 
These equations imply that the matter conservation equation
\be
\dot{\rho} + 3H\, (\rho + P) = 0\, ,
\ee
has the same form as in the classical theory, even when both Friedman and Raychaudhuri equations suffer loop quantum corrections. From Eq.~(\ref{dotH}) one sees that, for matter satisfying the WEC, there is a super-inflationary phase, corresponding to $\dot{H}>0$, whenever the matter density satisfies $\rho>\rcr/2$.
Note that in the $\lambda\to 0$ limit, we recover the corresponding classical equations.

Another system of interest, for the remainder sections of this Chapter, is a scalar field subject to a potential ${\rm V}(\phi)$. Even for the simplest potential $\v(\phi)=m^2\phi^2/2$ the classical dynamics is drastically modified; after the big bang there is a, `slow roll', inflationary period. A pressing question is how this dynamics gets modified in the effective LQC scenario. We know that every trajectory follows the effective Friedman equation (\ref{eff-fried}) and has a bounce when
$\rho=\rcr$, followed by a period of superinflation. How does that behavior affect the presumed
inflationary period occurring at much smaller densities? First note that in that case, the energy
density has the form: $\rho=\dot{\phi}^2/2 + m^2\phi^2/2$, so
there is a convenient way of depicting the bounce as the curve, in the $(\phi,\,\dot{\phi})$ plane, satisfying $\rcr=\dot{\phi}^2/2 + m^2\phi^2/2$. The dynamics is therefore bounded by such ellipsoid.
The equation satisfied by the scalar field has the same form as in the classical case: $\ddot{\phi} + 3H + \v_{,\phi}=0$. One can solve these equations numerically \cite{svv,ck-inflation} and finds that after the superinflationary phase, the dynamics follows very closely the GR dynamics and exhibits an  `attractor' behavior as well. As we shall see in later sections, this feature of the dynamics is responsible for phenomenologically relevant inflation to be generic.

Let us end this part with some comments. i) This set of effective equations have the property that
one recovers General Relativity in the small density `IR' limit, and that they are independent of the fiducial ${\cal V}$. These are non-trivial requirements that impose strong conditions on the particular form of the quantum constraint operator \cite{cs1}. ii) Inverse volume effects can introduce modifications to the effective equations that have various consequences, such as loss of the universal conservation equation for matter, and extra superinflationary corrections. However, the physical validity of considering such inverse correction for the $k$=0 is seriously challenged.
iii) It has been shown that for generic matter content, the LQC effective equations imply that
strong singularities are generically resolved \cite{ps}. iv) A consistency check for the
validity of effective equations pertains to the behavior of appropriately defined semiclassical states. Such states have been constructed and the predictions of the effective theory put to the test
\cite{cm1}. It was shown that both the density at the bounce and the minimum value of volume are very well described by the effective theory.

\subsubsection{$k$=1 FLRW}
\label{sec:2.c.2}

Let us now start with the isotropic closed FLRW model. As discussed before, there are two quantization available for this model. Correspondingly, the effective equations will yield two inequivalent theories.
For the first quantization, based in  the curvature as defined by closed holonomies, and neglecting the so called inverse triad corrections, one can arrive at the form of the effective Hamiltonian constraint,
\be
\mathcal{C}_{\textrm{eff}}=-\frac{3}{8\pi G\gamma^2\lambda^2}V\left[\sin^2(\lambda\b - D)-\sin^2D+(1+\gamma^2)D^2\right]+\rho V
\ee
with $D:=\lambda\vartheta/V^{1/3}$.
We can now compute the equations of motion from the effective Hamiltonian as,
$$\dot{V}=\{V,\mathcal{C}_{\textrm{eff}}\}=\{V,\b\}\frac{\partial\mathcal{C}_{\textrm{eff}}}{\partial\b}=\frac{3}{\lambda\gamma}V\sin(\lambda\b - D)\cos(\lambda\b - D)\, .
$$
From here, we can find the expansion as,
\be
\theta=\frac{\dot{V}}{V}=\frac{3}{\lambda\gamma}\sin(\lambda\b - D)\cos(\lambda\b - D)=\frac{3}{2\lambda\gamma}\sin2(\lambda\b-D)\label{exp-1}\, .
\ee
>From the above equation we can see that the Hubble parameter is also absolutely bounded by $|H|=|\theta|/3\leq 1/2\lambda\gamma$. We can now compute the modified, {\it effective Friedman equation}, by computing $H^2=\frac{\theta^2}{9}$,
\be
\begin{split}
H^2 & = \frac{1}{\lambda^2\gamma^2}\left(\frac{8\pi G\gamma^2\lambda^2}{3}\rho+\sin^2D-(1+\gamma^2)D^2\right)
\left(1-\frac{8\pi G\gamma^2\lambda^2}{3}\rho-\sin^2D+(1+\gamma^2)D^2\right)\\
&=\frac{8\pi G}{3}(\rho-\rho_1)\left(1-\frac{\rho-\rho_1}{\rcr}\right)
\end{split}\label{eff-frid-1}
\ee
where
$\rho_1=\rcr [(1+\gamma^2)D^2-\sin^2D]$ and
$\rcr=3/(8\pi G\gamma^2\lambda^2)$ is the  {\it critical density} of the $k=0$ FLRW model.

Let us now consider the other quantization, based on defining the connection using holonomies along open paths. As mentioned before, this is the only available route for anisotropic cosmologies when there is intrinsic curvature (such as in Bianchi II and IX). The effective Hamiltonian constraint one obtains from that quantum theory \cite{ck2}, when neglecting inverse scale factor effects (as was done in \cite{apsv} and \cite{ps-fv}), is
\be
\mathcal{C}_{\textrm{eff}}=-\frac{3}{8\pi
G\gamma^2\lambda^2}V\left[(\sin\lambda\b - D)^2+\gamma^2 D^2\right]+\rho V\, .
\ee
It is then straightforward to compute the corresponding effective equations of motion.
In particular, by computing $\dot{V}=\{V,\mathcal{C}_{\textrm{eff}}\}$, we can find the expression
for the expansion as
\be
\theta=\frac{3}{\lambda\gamma}\cos\lambda\b \left(\sin\lambda\b - D\right)\label{exp-2}\, .
\ee
Note that in this case, the expansion (and Hubble) is not absolutely bounded, due to the 
presence of the term linear in $D$. 
An important feature of these effective equations is that they describe with great accuracy the expectation value of volume during the numerical evolution of semiclassical quantum states \cite{apsv}. It is also worth notice that for large values of the recollapse volume, the effective and the classical equations coincide. In the case of the connection based quantization \cite{ck2}, there are two different bounces, that approach the unique bounce of the curvature based equations when the universe grows to be large \cite{ck2}. Let us now consider the effective equations for anisotropic models.

\subsubsection{Anisotropic Models: Bianchi I, II and IX}
\label{sec:2.c.3}

Considering the effective description of anisotropic models is interesting in view of the BKL conjecture \cite{bkl1,ahs}, that states that locally, generic spacetimes approaching the classical singularity behave as a combination of Bianchi cosmological models.
The effective Hamiltonian constraint for Bianchi I and II can be written in a single expression \cite{awe2,awe3,CKM},
\begin{align*} 
\label{H-BII}
\mathcal{C}_{\rm BII} & = \f{p_1p_2p_3}{8\pi G\gamma^2\lambda^2}
\left[\f{}{}\sin\bar\mu_1c_1\sin\bar\mu_2c_2+\sin\bar\mu_2c_2
\sin\bar\mu_3c_3+\sin\bar\mu_3c_3\sin\bar\mu_1c_1\right] \nonumber\\ 
& \quad + \f{1}{8\pi G\gamma^2}
\Bigg[\f{\alpha(p_2p_3)^{3/2}}{\lambda\sqrt{p_1}}\sin\bar\mu_1c_1 
-(1+\gamma^2)\left(\f{\varepsilon p_2p_3}{2p_1}\right)^2 \Bigg] - \f{p_\phi^2}{2} \approx 0 \, 
\end{align*}
where the parameter $\varepsilon$ allows us to distinguish between Bianchi I ($\varepsilon$= 0) and Bianchi II ($\varepsilon$= 1). 
This Hamiltonian together with the Poisson Brackets $\{c^i,p_j\}=8\pi G\gamma\delta_j^i$ and $\{\phi,p_\phi\}=1$
gives the effective equations of motion.
In these previous effective Hamiltonians, one chooses the lapse $N=V$.

In Bianchi IX, we choose $N$=1 
to include more inverse triad corrections, 
then the effective Hamiltonian is given by \cite{CKM}
\begin{equation*}
\label{H-BIX}
\begin{split}
\mathcal{C}_{\rm BIX}=&-\frac{V^4A(V)h^6(V)}{8\pi GV_c^6\gamma^2\lambda^{2}}\bigg(\sin\bar\mu_1c_1\sin\bar\mu_2c_2+\sin\bar\mu_1c_1\sin\bar\mu_3c_3\\
&+\sin\bar\mu_2c_2\sin\bar\mu_3c_3\bigg)
+\frac{\vartheta A(V)h^4(V)}{4\pi GV_c^4\gamma^2\lambda}\bigg(p_1^2p_2^2\sin\bar\mu_3c_3+p_2^2p_3^2\sin\bar\mu_1c_1\\
&+p_1^2p_3^2\sin\bar\mu_2c_2\bigg)
-\frac{\vartheta^2(1+\gamma^2)A(V)h^4(V)}{8\pi GV_c^4\gamma^2}\bigg(2V\bigg[p_1^2+p_2^2+p_3^2\bigg]\\
&-\bigg[(p_1p_2)^{4}+(p_1p_3)^{4}
+(p_2p_3)^{4}\bigg]\frac{h^6(V)}{V_c^6}\bigg)
+\f{h^6(V)V^2}{2V^6_c}p_\phi^2 \approx 0
\end{split}
\end{equation*}

Let us discuss the issue of singularity resolution when these equations are studied numerically. 
i) All solutions have a bounce. In other words, singularities are resolved.
In the closed FRW and the Bianchi IX model, there are infinite number of bounces and recollapses due to the compactness of the spatial manifold. ii) 
One can have a different kind of bounce dominated by shear $\sigma$, but only in Bianchi II and IX. In Bianchi I, the dynamical
contribution from matter is always bigger than the one from the shear, even in the solution which reaches the maximal shear at the bounce \cite{ac-bianchi}. iii)
In the flat isotropic model all the solutions to the effective equations
have a maximal density equal to the critical density, 
and a maximal expansion ($\theta^2_{\rm max} = 6\pi G \rcr = 3/(2\gamma\lambda)$) 
when $\rho=\rho_{\rm crit}/2$. 
For FRW $k=1$ model, every solution has its maximum density but in general the density is not absolutely bounded. In the effective theory which comes from connection based quantization, expansion can tend to infinity. For the other case, expansion has the same bound as the flat FRW model. However, by adding some more corrections from inverse triad term, one can show that actually in both effective theories the density and the expansion have finite values.
iv) For Bianchi I, in all the solutions $\rho$ and $\theta$ are upperly bounded by its values in the isotropic case
and $\sigma$ is bounded by $\sigma^2_{\max} = 10.125/(3\gamma^2\lambda^2)$ \cite{ps-bianchi,singh-gupt}. For Bianchi II, $\theta, \sigma$ and $\rho$ are also bounded, but for larger values than the ones in Bianchi I,  i.e., there are solutions where the matter density is larger than the
critical density. With point-like and cigar-like classical singularities \cite{ac-bianchi},
the density can achieve the maximal value ($\rho \approx 0.54\rho_{\rm Pl} $) as a consequence of the shear being zero at the bounce and curvature different from zero.
v) For Bianchi IX the behaviour is the same as in closed FRW, if the inverse triad corrections are not used, then the geometric scalars are not absolutely bounded. But if the inverse triad corrections are used then, on each solution, the geometric scalars are bounded but there is not an absolute bound for all the solutions \cite{CKM,singh-gupt}.

\section{Inhomogeneous perturbations in LQC}
\label{sec:3}
 
The theory of quantized fields in curved space-times has become an essential tool in modern  early-universe cosmology. In that framework one studies the behavior of quantum fields propagating in space-times with generic Lorentzian geometries, as in General Relativity.  One expects this theory to describe accurately physical processes in situations where we are confident about the validity of its building blocks: a description of matter fields in terms of quantum field theories, and a space-time geometry given by a smooth, classical space-time metric. These assumptions are reasonable, for instance, during the inflationary era in which the energy density and curvature are believed to be more than ten orders of magnitude below the Planck scale. However, earlier in the history of the universe, closer to the Planck era, quantum gravity effects become important and the description of space-time geometry in terms of a smooth metric is expected to fail. To include physics in the Planck regime QFT in curved backgrounds needs to be generalized to a QFT in {\em quantum} space-times. The  singularity-free quantum geometry provided by LQC,  summarized in the previous section, provides a suitable arena to formulate such a theory, and the quantization of scalar fields on those quantum cosmologies was introduced by Ashtekar, Kaminski and Lewandowski in \cite{akl}, and further developed in \cite{aan2,puchta, dapor1,dapor2}.  Having in mind the most interesting application of this framework, we summarize here the construction of the QFT of scalar and tensor metric perturbations propagating in a quantum FLRW universe, i.e. the {\em quantum gravity theory of cosmological perturbations}. For more detailed information, see \cite{akl, aan2}.

As mentioned in the introduction of this chapter, the construction will follow the guiding principle that has been useful in the quantization of the background: first carry out a truncation of the classical theory to select the sector of General Relativity of interest, and then move to the quantum theory by using LQG techniques. Starting from General Relativity with a scalar field as matter source, we will truncate the phase space to the sector containing cosmological backgrounds {\em plus} inhomogeneous, gauge invariant, first order perturbations, and then write down the dynamical equations on that classical, reduced phase space. The main approximation behind this truncation, and underlaying the subsequent quantization, is that  the back-reaction of inhomogeneous perturbations on the homogeneous degrees of freedom is neglected. The second step is to move to the quantum theory. Physical states will depend on background homogeneous degrees of freedom as well as on inhomogeneous ones. Our basic approximation, however, enables us to write these quantum states as tensor product of the homogeneous part, which will evolve independently of perturbations, and first order inhomogeneities thereon. The homogeneous part will therefore be the same as the quantum geometries obtained in the previous section, in which the big bang singularity is replaced by a bounce. The surprising result appears in the evolution of perturbations. Without further approximation, the evolution of inhomogeneities on those quantum geometries turns out to be {\em mathematically equivalent} to the quantum theory of those fields propagating on a {\em smooth} background characterized by a metric tensor. The components of that smooth metric, however, do not satisfy the classical Einstein equation. They are obtained from expectation values of certain combinations of background operators, and incorporate {\em all} the information of the underlying quantum geometry that is `seen' by perturbations. The message  is that the propagation of inhomogeneous perturbations is not sensitive to all the details of the quantum space-time, but only to certain aspects, which appear precisely in a way that allows to encode them in a smooth background metric. This is an unforeseen simplification that facilitates enormously the treatment of field theoretical issues.

The last step is to develop the necessary tools to check the self-consistency of this construction. It is necessary to show that, in the physical situations under consideration, the Hilbert space of physical interest contains a large enough subspace in which the back-reaction of perturbations on the background is indeed negligible, in such a way that our initial truncation is justified. That should be done by comparing the expectation value of the Hamiltonian and stress-energy tensor for perturbations with that of background fields. Those computation will require of techniques of regularization and renormalization.\\

\subsection{The classical framework}
\label{sec:3.a}

The goal of this subsection is to summarize the construction of the truncated theory of classical FLRW space-times coupled to a scalar field,  plus gauge invariant, linear perturbations on it, and write down the equations describing their dynamics. The reader is referred to the extensive literature for more details  (see, for instance, \cite{reportbrandenberger}). We adopt here the Hamiltonian framework which, as shown in \cite{langlois}, is particularly transparent on the task of finding gauge invariant variables. It will also provide the appropriate arena to pass to the quantum theory in the next section. For simplicity and for physical interest, we work here with a spatially flat FLRW universe. 

The procedure can be divided in three steps: 1) Starting from the full phase space, expand the configuration variables and their conjugate momenta in perturbations, and truncate the expansion at first order. Expand also the constraints of the theory (the scalar and vector constraints) and keep only terms containing zero and first order perturbations. 2) Use the constraints linear in first order perturbations to find gauge invariant variables. Those variables coordinatize the so-called truncated, reduced phase space. 3) Use the part of the constraints quadratic in zero and first order perturbations to write down the dynamics. See \cite{aan2} for further details and subtle points of this construction.

\subsubsection{The truncated phase space} 
\label{sec:3.a.1}

Let us consider General Relativity coupled to  a scalar field on a space-time manifold $M=\Sigma\times \mathbb{R}$, with $\Sigma=\mathbb{R}^3$. Due to the infinite volume in $\Sigma$, spatial integrals of homogeneous quantities will introduce infrared divergences. To be able to write meaningful mathematical expression, it is convenient to introduce a fiducial cell ${\cal V}$ and restrict all integrals to it.  ${\cal V}$ can be chosen to be arbitrarily large, or at least larger than the observable universe. At the quantum level this will be equivalent to restrict to ${\cal V}$ the support of test functions in operator valued distributions.

We will work with ADM variables for the gravitational sector, where the canonical conjugated pairs consist in a  positive definite 3-metric on $\Sigma$, $q_{ab}$, and its conjugate momentum $p^{ab}$ (the same analysis can be done in connexion variables, by including  the corresponding Gauss constraint; see \cite{dt,pert_scalar,ghtw,joao2}, \cite{aan2}). The full phase space $\Gamma$ consists of quadruples $\{q_{ab}(\vec{x}),p^{ab}(\vec{x}),\Phi(\vec{x}),\Pi(\vec{x})\}\in\Gamma$, where $\Pi(\vec{x})$ is the conjugate momentum of the scalar field $\Phi(\vec{x})$. Because we are interested in expanding around $\Gamma_{\rm hom} \subset \Gamma$, the (FLRW) isotropic and homogenous sector of $\Gamma$, it is convenient to introduce a fiducial flat metric $\qzero_{ab}$, and use it to raise and lower indices. We will denote $\vec{x}=(x_1,x_2,x_3)$ the Cartesian coordinates defined by $\qzero_{ab}$ on ${\cal V}$, $\Vzero$ the volume of ${\cal V}$ with respect to $\qzero_{ab}$, which we take equal to one to simplify the notation, and $\qzero=1$ the determinant of $\qzero_{ab}$.

Consider now curves $\gamma[\ep]$ in $\Gamma$, which pass through $\ps_{\hom}$ at $\ep=0$. Expanding the phase space variables around $\ep=0$, we have: 
\ba \label{expan} q_{ab}[\ep](\x) &=& a^2 \qzero_{ab} + \ep\,
\delta q^{(1)}_{ab}(\x) + \ldots + \f{\ep^n}{n!}\, 
\delta q^{(n)}_{ab}(\x) + \ldots
\nonumber\\
p^{ab}[\ep](\x) &=& \, \frac{P_{a}}{6\,  a} \qzero^{ab} + \ep\, \delta p^{ab\, (1)}(\x) + \ldots \, \nonumber\\
\Phi[\ep](\x) &=& \phi + \ep\, \vpone(\x) + \ldots \, , \nonumber \\
\Pi[\ep](\x) &=& \, \pphi + \ep\, \pione(\x)+ \ldots 
 \ea
It is convenient to consider  the first order perturbations $\delta q^{(1)}_{ab}(\x), \, \delta p^{ ab \, (1)}(\x), \vpone(\x), \pione(\x)$ as \emph{purely
inhomogeneous} functions of $\x$, in the sense that the integral of any of them on ${\cal V}$ is zero. By truncating the above expansions at first order we obtain the {\em truncated} phase space, made of four pairs of conjugate variables: $\Gamma_{\rm{Trun}}=\{(a,P_{a},\phi, \pphi,\delta q^{(1)}_{ab}, \, \delta p^{ ab \, (1)}, \vpone, \pione)\}=\Gamma_{\rm hom}\times\Gamma_1$, where the only non-zero Poisson brackets between the basic variables are:
\ba \label{pbs}\{a,\, P_{a}\} = 1, &\quad& \{\delta q^{(1)}_{ab}(\x_1),\,  \delta p^{ cd \, (1)}(x_2)\} = \delta^c_{(a}\, \delta_{b)}^d\, \bar\delta(\x_1,\x_2),\nonumber\\ \{\phi,\, \pphi\} = 1, &\quad& \{\vpone(\x_1),\, \pione(\x_2)\} = \bar\delta(\x_1, \x_2), \ea
where $\bar\delta(\x_1,\x_2) = (\delta(\x_1,\x_2) - 1)$ is the Dirac delta distribution on the space of purely inhomogeneous fields.  
>From now on we will work only with first order perturbations, so we will omit the super-index $(1)$ to simplify the notation.

Because of the homogeneity of the background it is convenient to Fourier transform the perturbation fields and carry out the standard scalar-vector-tensor decomposition, in which the 6 degrees of freedom of $\delta q_{ab}$ are decompose into two scalar, two vector, and two tensor modes (see e.g.\cite{langlois}, \cite{aan2} for details). Because perturbations are inhomogeneous, the restriction to the fiducial cell $\cal{V}$ is not strictly necessary, and one can avoid the artificial quantization of $\vec{k}$ that it introduces. However,  from the physical point of view one can absorb modes with wavelength larger than the observable universe in the background. Therefore, we will consider that the Fourier integrals incorporate an infrared cut-off $k_o$ provided by the size of the observable universe.

\subsubsection{Constraints and reduced phase space}
\label{sec:3.a.2}

A similar expansion to (\ref{expan}) can  be carried out for the constraints. In General Relativity the Hamiltonian is a sum of constrains, the familiar scalar $\mathbb{S}[N]$, and vector $\mathbb{V}[\vec{N}]$ constraints. If $\gamma[\epsilon]$ is now a curve that lies in the constraint hyersurface of $\Gamma$, and intersects $\ps_{\hom}$ at $\ep=0$, by referring to the constraints collectively as ${\cal C}(q^{ab},p_{ab}, \Phi,\Pi)$ (suppressing the smearing fields for simplicity), we expand around $\epsilon=0$ to obtain a hierarchy of equations:
\be {\cal C}^{(0)}:={\cal C}|_{\ep =0}= 0,\quad {\cal C}^{(1)}:=\f{d {\cal C}}{d\ep}|_{\ep=0} = 0, \quad
\ldots\quad {\cal C}^{(n)}:=\f{d^n {\cal C}}{d\ep^n}|_{\ep=0} = 0,\quad \ldots \ee

\begin{itemize} 
\item
The zeroth-order constraint, ${\cal C}^{(0)}= 0$, is just the restriction of the full constraint to the homogeneous subspace $\ps_{\hom}$. The zeroth-order vector constraint is trivially satisfied because of the gauge fixing on the zeroth-order variables, introduced by the use of the fiducial metric $\qzero_{ab}$ in (\ref{expan}). The zeroth-order scalar constraint, $ {\mathbb{S}}_{0}$, is quadratic in zeroth-order variables and can be interpreted as the generator of  background dynamics. This dynamics is exactly the same as that of the unperturbed theory.

\item First order constraints are linear in first order variables. They generate gauge transformations in $\Gamma_{\rm Trun}$ and, as usual, tell us that some of our degrees of freedom are not physical. Initially we have $6\, (\times \infty)$  degrees of freedoms in $\delta q_{ab}(\x)$, plus 1 degree of freedom in the scalar field $\varphi(\vec{x})$, a total of 7. As mentioned above, $\delta q_{ab}(\x)$ is conveniently decomposed in Fourier space into two scalars, two vector, and two tensor modes. 
We have the scalar and three vector constraints, a total of 4. Therefore, the number of physical degrees of freedom is $7-4 =3$. There is an elegant systematic procedure to construct  gauge invariant variables out of those 3 degrees of freedom, and we refer the reader to \cite{langlois} for details. It can be summarize as follows. In FLRW backgrounds, scalar perturbations are affected by the scalar constraint and only one of the vector constraints; they reduce the three scalar degrees of freedom that we have initially, two from gravity and one from the matter sector, to only one physical scalar mode. Vector perturbations are affected by two of the vector constraints that kill completely the vector modes. In other words, in the absence of matter with vector degrees of freedom, as in the case we are studying, there are no physical vector perturbations. Tensor modes are not affected by any of the constraints and therefore the two original tensor modes are the physical ones, i.e. they are gauge invariant. In summary, after imposing the constraints we are left with one scalar degree of  freedom, which we choose to be the familiar Mukhanov variable $\Q$, and two tensor modes $\T^{(1)}$ and $\T^{(2)}$. They are gauge invariant variables. and together with their conjugate momenta form the {\em reduced}, truncated phase space of first-order perturbations, $\tilde{\Gamma}_{\rm Trun}$.  Equations ${\cal C}^{(n)}=0$ with $n>1$ do not add further constraints on first oder perturbations.

\item
The second-order constraints in the full phase space $\Gamma$ involve terms quadratic in first-order perturbations as well as linear terms in second-order perturbations. When a second order constraint ${\cal C}^{(2)}$ is restricted to the truncated phase space $\tilde\Gamma_{\rm Trun}$, terms containing second order perturbations are disregarded, and the resulting combination of quadratic terms in first-order perturbation with coefficients containing  background quantities,  $\tilde {\cal C}^{(2)}$, {\em is no longer a constraint}. The truncated second-order scalar constraint   $\tilde {\mathbb{S}}_{2}$ is interpreted as the Hamiltonian that generates the dynamics of gauge invariant first-order perturbations. It  has the form $\tilde {\mathbb{S}}_{2}=\tilde {\mathbb{S}}^{(\Q)}_{2}+\tilde {\mathbb{S}}^{(\T^{(1)})}_{2}+\tilde {\mathbb{S}}^{(\T^{(2)})}_{2}$,  which indicates that scalar and tensor modes evolve independently of each other, where

\be \label{pert-ham} \tilde{\mathbb{S}}_2^{(\T)}[N]= \f{N}{2 (2\pi)^3 }\,\, \int d^3 k \, \left( \f{4 \kappa}{a^{3}}\, |\pp^{(\T)}_{\vk}|^2 + \frac{a\, k^2}{4 \kappa} |\T_{\vk}|^2 \right)\, . \ee
%
%
with $\kappa=8\pi G$. The two tensor modes behave identically, and we have denoted them collectively by $\T$. For pedagogical reasons we only write down the expressions for tensor perturbations. See \cite{aan2,aan3} for explicit expressions for scalar modes.  In the above equations $\pp^{(\T)}_{\vk}$ 
 are the conjugate momenta of $\T_{\vk}$,  
 with Poisson brackets $\{ \T_{\vk},p^{(\T)}_{-\vk'} \}=(2\pi)^3\delta(\vk-\vk')$. 
 Tensor perturbations, except for the constant factor $1/(2\sqrt{\kappa})$ that provides the appropriate dimensions, behave exactly as massless, free scalar fields (scalar perturbations $\Q_{\vec{k}}$ behave as a scalar field subject to a time dependent `emergent' potential).  
  The (homogeneous) lapse function $N$ indicates the time coordinate one is using. For instance, $N=1$ corresponds to standard cosmic time $t$, $N=a$ to conformal time $\eta$, and $N=a^3 /\pphi$ to choosing the scalar field $\phi$ as a time variable, which turns out to be the natural choice in the quantum theory.

\end{itemize}

To summarize, the phase space of  physical interest is the reduced, truncated phase space $\tilde\Gamma_{\tr}$ made of elements $ \{ (a,P_{a},\phi, p_{(\phi)}); (\Q_{\vec{k}},\pp^{(\Q)}_{\vk},\T^{(1)}_{\vk}, \pp^{(\T^{(1)})}_{\vk},\T^{(2)}_{\vk}, \pp^{(\T^{(2)})}_{\vk})\}\in \tilde\Gamma_{\tr}$. The homogenous degrees of freedom evolve with the zeroth-order Hamiltonian. This evolution takes place entirely in $\Gamma_{\rm hom}$, and is independent of perturbations, reflecting the main approximation of the truncated theory. The homogenous dynamical trajectory can  then be `lifted' to $\tilde\Gamma_{\tr}$, providing a well-defined evolution of first-order perturbations on the homogenous background. This evolution is specified by the Hamiltonian $\tilde {\mathbb{S}}_{2}$.

\subsection{Quantum theory of cosmological perturbations on a quantum FLRW \label{QFTQST}}

\subsubsection{Quantization of  $\tilde \Gamma_{\rm Trunc}$}
\label{3.b.1}

In this section we pass to the quantum theory starting from the reduced, truncated phase space $\tilde\Gamma_{\tr}$. 
The structure of the classical phase space $\tilde\Gamma_{\tr}=\Gamma_{\rm hom}\times \tilde \Gamma_{1}$  suggests that in the quantum theory the total wave function $\Psi$ has the form

\be \label{tenpro}  \Psi(a,\T_{\vk}, \phi)=\Psi_0(a,\phi)\otimes \psi(\T_{\vk}, \phi) \, . \ee
This product structure is maintained as long as the test field approximation holds. Because back-reaction is neglected, the background part $\Psi_0$ evolves independently of perturbations, and the solutions  for $\Psi_0$ are the ones obtained in section \ref{sec:2}. When  written in terms of the relational time $\phi$, they satisfy the equation
$\hat{p}_{(\phi)} \Psi_o \equiv -i\hbar\, \partial_\phi \Psi_0 = \h{H}_0\Psi_0$,  where the operator $\h{H}_0\equiv \sqrt{\Theta}$ is obtained from expressions (\ref{hc5}) and (\ref{schr-eq}).  The remaining task is to `lift' this trajectory to the full Hilbert space, by writing down the quantum theory for $\psi$ propagating on the quantum geometry specified by $\Psi_0$. The evolution of $\psi$ will be specified by the  operator analogue of $\tilde {\mathbb{S}}_{2}^{(\T)}$, which generates the dynamics in the classical phase space. In the classical theory $\tilde {\mathbb{S}}_{2}^{(\T)}$  depends on inhomogeneous degrees of freedoms, but also on the homogeneous ones via the scale factor $a$. Therefore, in the quantum theory the corresponding operator will act  on perturbations $\psi$ as well as on $\Psi_0$. 

Our goal is to generalize the theory of QFT in  curved space-times in which, on the one hand, quantum fields propagate in an {\em evolving} classical FLRW specified by $a_{\rm cl}(\eta)$ and, on the other hand, perturbations are commonly quantized using the Heisenberg picture. Therefore, to facilitate the comparison, we pass in this section to the Heisenberg picture. In obtaining the evolution equations for the operator $\hat\T_{\vk}$ and its  conjugated momentum we will use  $\phi$ as internal time, because it is the evolution variable that appears naturally in the quantum theory, while standard cosmic or conformal time are represented by operators. Internal time $\phi$ corresponds to use the lapse function $N =a^3 /\pphi$ in the expression (\ref{pert-ham}). By choosing an appropriate factor ordering to convert it to an operator, we have (as it is common in quantum theory, we are not free of factor ordering ambiguities) 

\ba  \label{eqmotop}   \partial_{\phi} \h\T_{\vk}(\phi) =\frac{i}{\hbar}[\h\T_{\vk}, \h{\tilde{\mathbb{S}}}_{2}^{(\T)}]&=&  \, 4 \kappa \, (\hat{p}_{(\phi)}^{-1}\otimes \h{\pp}^{(\T)}_{\vk} ) \, ; \nonumber \\ \partial_{\phi} \h\pp^{(\T)}_{\vk}(\phi) =\frac{i}{\hbar} [ \h \pp^{(\T)}_{\vk},\h{\tilde{\mathbb{S}}}_{2}^{(\T)}]&=&- \, \frac{ k^2}{4 \kappa} \, ( \hat{p}_{(\phi)}^{-1/2} \,\h a^4(\phi) \, \hat{p}_{(\phi)}^{-1/2} \otimes \h \T_{\vk} ) \, .\ea
These equations involve background operators as well as perturbations. However, the test field approximation allows us to `trace over' the background degrees of freedom. This can be done by taking expectation value with respect to the background wave function $\Psi_0$ (in the Heisenberg picture)  obtained in the previous section

\ba \label{eqmot} \partial_{\phi} \h\T_{\vk}&=&  \, 4 \kappa \, \langle \h H_0^{-1}\rangle \, \h\pp^{(\T)}_{\vk}  \, , \nonumber \\
\partial_{\phi} \h\pp^{(\T)}_{\vk} &=& - \, \frac{ k^2}{4 \kappa} \, \langle \h H_0^{-1/2} \, \h a^4(\phi)\, \h H_0^{-1/2} \rangle \,  \h \T_{\vk}   \, , \ea
where background operators have been replaced by expectation values and, additionally,  we have used the evolution equation $\hat{p}_{(\phi)} \Psi_o  = \h{H}_0\Psi_0$. The test field approximation ensures that we are not losing any information when passing from (\ref{eqmotop}) to (\ref{eqmot}).
These are the Heisenberg equations for perturbations, in which the coefficients are given by {\em expectation values of background operators in the quantum geometry specified by} $\Psi_0$. This is a quantum field theory of cosmological perturbation on a {\em quantum} FLRW universe. Note that the above equation is exact, and not further approximation has been made beyond the test field approximation. 

In this theory, space-time geometry is no described by a unique classical metric,  it is rather characterized by a probability distribution $\Psi_0$ that contains the unavoidable quantum fluctuations. The propagation of perturbations is sensitive to those fluctuations, and not only the mean effective trajectory $\langle \h a \rangle$. However, it is remarkable that those effects can be encoded in a couple of  expectation values of background operators: $\langle \hat{H}_o^{-1}\rangle$ and $\langle \hat{H}_o^{-\f{1}{2}}\,\hat{a}^4(\phi)\, \hat{H}_o^{-\f{1}{2}}\rangle$ \cite{akl, aan2}.  
Borrowing the analogy from \cite{aan2}, this is similar to what happens in the propagation of light in a medium: the electromagnetic waves interact in a complex way with the atoms in the medium, but the net effect of those interactions can be codified in a few parameters, such as the refractive index. Similarly, although the final equations (\ref{eqmot}) depend in a simple way on the quantum geometry, it had be very difficult to guess the precise `moments' of the quantum geometry that are involved in the evolution of perturbations. 

We can now compare the above evolution equations with the familiar quantum field theory of cosmological perturbations on classical FLRW geometries, in which the Heisenberg equations, when $\phi$ is used as time, are written in terms of the classical background quantities $a(\phi)$ and $p_{(\phi)}$ as

\be \label{claseqmot} \partial_{\phi} \h\T_{\vk} =  \, \frac{4 \kappa}{p_{(\phi)}} \, \h\pp^{(\T)}_{\vk}  \, ; \quad \quad  \partial_{\phi} \h\pp^{(\T)}_{\vk} =- \, \frac{k^2}{4 \kappa} \, \frac{a(\phi)^4}{ p_{(\phi)}}  \, \h \T_{\vk}   \, .\ee
Comparing with ({\ref{eqmot}) we see that the QFT in a quantum background $\Psi_o$ is {\em indistinguishable} from a QFT on a {\em smooth FLRW metric }
\be \tilde{g}_{ab}\, \d x^a \d x^b \equiv \d\tilde{s}^2 = -
(\tilde{p}_{(\phi)})^{-2}\, \tilde{a}^6(\phi)\, \d\phi^2 +
\tilde{a}(\phi)^2\, \d \vec{\mathrm{x}}^2 \ee
where
\be (\tilde{p}_{(\phi)})^{-1}  = \langle \hat{H}_o^{-1}\rangle
\quad\quad {\rm and} \quad\quad \tilde{a}^4 = \f{\langle
\hat{H}_o^{-\f{1}{2}}\, \hat{a}^4(\phi)\,
\hat{H}_o^{-\f{1}{2}}\rangle}{\langle \hat{H}_o^{-1}\rangle}\, .
\ee
In terms of the more familiar conformal time used in cosmology, we have $\d\tilde{s}^2 = \tilde{a}^2(\tilde\eta)\, (-\d\tilde\eta^2 +
\, \d\vec{x}^2)$, with $\d\tilde{\eta} = [ \tilde{a}^2(\phi)]\,
\tilde{p}_{(\phi)}^{-1}\, \d\phi$. This smooth metric captures all the information of quantum geometry that is `seen' by perturbations. Note that its components contain $\hbar$ and it does not satisfy the Einstein equation, not even the LQC effective equations. 

In terms of this smooth metric, we can write the Heisenberg equations (\ref{eqmot}) as a second order differential equation
\be \label{Teqn} \hat{\T}_{\vk}^{\prime\prime} + 2
\f{\tilde{a}^\prime}{\tilde{a}}\, \hat{\T}_{\vk}^\prime + k^2
\hat{\T}_{\vk} = 0 \, ,\ee
where the prime now denotes derivative with respect to $\tilde\eta$. 
This equation is mathematically equivalent to the familiar formulation of QFT in classical FLRW space-time, where all the effects of the quantum background geometry have been encoded in a {\em dressed, smooth metric tensor} $\tilde g_{ab}$. This unexpected mathematical analogy highly simplifies the analysis, not only conceptually, but also at the technical level. It allows to extend well-stablished techniques from classical space-times to define the physical Hilbert space and the appropriate regularization and renormalization of composite operators on it (see \cite{aan2,aan3} for details of that construction). These are the necessary tools to make sense of the momentum integrals appearing in, e.g. the Hamiltonian $\h{\tilde{\mathbb{S}}}_2$, that so far were formal, and to regularize the expectation value of the energy-momentum tensor in the physical Hilbert space.

%
%
%

\subsubsection{The physical Hilbert space \label{hilbertspace}}

In this subsection we briefly summarize how techniques of regularization and renormalization from linear QFT on classical space-times, can be extended to characterize the physical Hilbert space of cosmological perturbations on quantum backgrounds, and to regularize composite operators on it. Among the existing methods of regularization we will work in the adiabatic approach \cite{parker66, parker-fulling74}, which is particularly convenient to perform explicit computations, including the numerical  implementation required in the next section.

The spatial homogeneity and isotropy of our FLRW background allows us to expand the field operator $\h \T(\x,\tilde\eta)$ in Fourier modes (a similar construction holds for scalar perturbations)

\be \label{fieldexp} \h \T(\x,\tilde\eta)=\frac{1}{(2\pi)^3} \int \d^3k \left( \h A_{\vk} \, e_k(\tilde\eta)+\h A^{\dagger}_{\vk} \, e^{\star}_k(\tilde\eta) \right) \, e^{i \vk\x} \, . \ee
The field operator $\h \T(\x,\tilde\eta)$ satisfies the equation of motion (\ref{Teqn}) as long as the mode functions $e_k(\tilde\eta)$ are solution of the wave equation
\be \label{we} e''_k(\tilde\eta)+2 \frac{\tilde a''}{\tilde a}\, e'_k(\tilde\eta)+k^2 \, e_k(\tilde\eta)=0 \, ,\ee
were prime indicates derivative with respect to $\tilde \eta$. The solutions $e_k(\tilde\eta)$ can be understood as `generalized positive frequency modes', because they play the role of standard positive frequency  solutions $e^{-i k t}/\sqrt{2 k}$ in Minkowski space-time. The canonical commutation relations for the field operator $\h \T(\x,\tilde\eta)$ and its conjugate momentum imply 
\be [\h A_{\vk} ,\h A^{\dagger}_{\vk'}] = i \hbar  \, (2\pi)^3  \, \delta(\vk-\vk')  \, \langle e_k(\tilde\eta),e_{k'}(\tilde\eta)\rangle^{-1}  \, ; \quad   [\h A_{\vk} ,\h A_{\vk'}]=0 \, , \ee
where
\be\langle e_k(\tilde\eta),e_{k'}(\tilde\eta)\rangle := \frac{ \tilde a^2}{4 \kappa}(e_k(\tilde\eta)e'^{\star}_{k'}(\tilde\eta)-e'_k(\tilde\eta)e^{\star}_{k'}(\tilde\eta)) \, . \ee
Therefore,  if we impose the normalization condition $\langle e_k(\tilde\eta),e_{k}(\tilde\eta)\rangle= i$, $\h A_{\vk}$ and $\h A^{\dagger}_{\vk}$ will satisfy the familiar commutation relations of creation and annihilation operators. Note that the scalar product $\langle e_k(\tilde\eta),e_{k'}(\tilde\eta)\rangle$ is constant in time if $e_k(\tilde\eta)$ and $e_{k'}(\tilde\eta)$ are solutions of (\ref{we}). The Hilbert space is then constructed as follows. The vacuum state $|0\rangle$ (associate with the set of generalized positive frequency modes $e_k$) is defined as the state annihilated by all $\h A_{\vk}$. The associated Fock space $\H_1$ arises by the repeated  action of creation operators $\h A^{\dagger}_{\vk}$ on the vacuum. It is important to notice that the vacuum state constructed in this way is {\em translational and rotational invariant}, as can be checked, e.g. by explicit construction of the two point function.

It is clear from the construction that a different choice for the generalized positive frequency bases $e_k$ in (\ref{fieldexp}), provides different $\h A_{\vk}$ and $\h A^{\dagger}_{\vk}$  operators, and therefore a {\em different definition of vacuum state}. None of those vacua is preferred as compare to the others. Even more, different vacua may not even belong to the same Hilbert space, and the quantum theories constructed from each of them are in that case unitarily inequivalent. 

The existence of unitarily inequivalent quantization is common in QFT in curved space-times (see e.g. \cite{waldbook}). In cosmological backgrounds, however, it is  possible to add appropriate regularity conditions to the mode functions $e_k$ to select a preferred Hilbert space. The {\em adiabatic condition} \cite{parker66,parker69,parker-fulling74}  in FLRW backgrounds imposes that, in the asymptotic limit in which the physical  momentum $k/\tilde a$ is much larger than the energy scale provided by the space-time curvature $E_R$, $e_k$ must approach the Minkowski space-time positive frequency modes, $e^{-i k t}/\sqrt{2 k}$, {\em at an appropriate rate} (for a brief summary see, e.g. \cite{aan2}, and references cited there). The modes $e_k$ satisfying this conditions are called  modes of $Nth$ adiabatic order, and the associated vacuum an adiabatic vacuum of the same order, where the  order is specified by the exact rate of approach to the Minkoswkian solutions. 

Notice that the adiabatic condition does not single out a preferred vacuum, because there are many different families $e_k$ satisfying it to a given order (it imposes only an {\em asymptotic} restriction for large $(k/\tilde a)/E_R)$. However, it is possible to show that if we restrict to adiabatic order $N\geq 2$, {\em all different vacua belong to the same Hilbert space} $\H_1$. (This is strictly true if we restrict our QFT to the compact fiducial cell $\cal V$. In the non-compact case one needs to be more precise in the sense in which the Hilbert space is unique \cite{waldbook}, because infra-red divergences appear). Additionally, if $N\geq 4$ there is a well defined procedure to extract the physical, finite information from the formal expression of operators of interest for us, the Hamiltonian and the stress-energy tensor, by subtracting ultra-violet divergences in a local and state independent way, while respecting the covariance of the theory.

The Hamiltonian  operator generating time evolution (in conformal time), and the energy density $\h \rho$, are related by 

\be \hat{ \tilde {\mathbb{S}}}^{(\T)}_{2,{\rm formal}}=\f{1}{(2\pi)^3 }\,\, \int \d^3 k \,  \frac{2 \kappa}{\tilde{a}^{2}}\, |\h \pp^{(\T)}_{\vk}|^2 + \frac{\tilde{a}^2 \, k^2}{8 \kappa} |\h \T_{\vk}|^2=\tilde a^4\int \d^3x \,  \h \rho^{(\T)}_{\rm formal}  \, . \ee
If $|0\rangle$ is a 4th-order adiabatic vacuum associated with a family of solutions $e_k(\tilde\eta)$, the renormalized expectation value of the energy density is given by
\be \langle 0|\h \rho^{(\T)}(\tilde\eta)| 0\rangle_{\rm ren} =\frac{\hbar}{8 \kappa \tilde{a}^2} \int \frac{\d^3k }{(2\pi)^3} \left[ |e'_k|^2+k^2 |e_k|^2 -\frac{4\kappa}{\tilde{a}^2} \,  C^{(\T)}(k,\tilde\eta)\right]\, .\ee
with
\be C^{(\T)}(k,\tilde\eta)=k+\frac{{\tilde{a}}'^2}{2 {\tilde{a}}^2 k}+\frac{4 {\tilde{a}}'^2
{\tilde{a}}''+\tilde{a} {\tilde{a}}''^2-2 \tilde{a} {\tilde{a}}'\, {\tilde{a}}^{'''}}{8
{\tilde{a}}^3 k^3} \, , \ee
where $C(k,\tilde\eta)$ are the  subtraction terms provided by adiabatic regularization \cite{parker-fulling74}. The renormalized expression for the expectation value of the  Hamiltonian operator is obtained from the previous three equations. The above subtractions make the expectation value of the hamiltonian and energy density finite for {\em any state} in the Hilbert space of 4th-order adiabatic states, $\H_1$. Additionally, the procedure has the properties that  any method of regularization/renormalization are expected to satisfy, enunciated in the Wald's axioms \cite{waldbook}. Although strictly speaking the above expressions provide only quadratic forms in the Hilbert space, recent results indicate that they are expectation values of operator value distributions $\h \rho^{(\T)}$ and $\tilde {\mathbb{S}}^{(\T)}_{2}$ in $\H_1$.



In summary, our QFT in quantum FLRW admits a straightforward extension of the adiabatic approach of linear QFT in classical backgrounds. The physical Hilbert space $\H_1$ is then singled out by restricting to 4th order adiabatic states. In addition, the adiabatic condition provides  the necessary control on ultra-violet divergences  that allows a systematic  procedure to regularize the Hamiltonian and the stress-energy tensor on $\H_1$. This completes the formulation of the theory. 


\subsubsection{Criterion for self-consistency\label{selfconsistency}}

The last step in the construction is to check whether the underlaying approximation in our truncated theory, the test field approximation, is satisfied throughout the evolution. In our QFT in quantum space-times this question translates to check whether the expectation value of the stress-energy tensor can be neglected when compared to the background one. However, in an homogeneous and isotropic background a sufficient  condition for this to be  satisfied is that energy density on scalar and tensor perturbations, $\langle \h \rho(\tilde\eta)\rangle$, be much smaller than the background energy density $\langle \rho_o \rangle $ {\em at any time} during dynamical phase of interest \cite{aan2}. It is evident that one can always find states for perturbation for which that  requirement is not satisfied. Therefore, the relevant question is: is there a sufficiently large subspace of the physical Hilbert space for which the previous condition on the energy density  is satisfied? If the answer is in the affirmative then one has a self-consistent approach in which test-field approximation holds. This is a key question to ensure self-consistency, and has to be answered when this framework is applied to a concrete physical problem, as we do in the next section.

\subsection{Comments}

The previous framework is suitable to face interesting conceptual questions arising in quantum gravity. For instance, when does standard QFT in curved space-times become a good approximation? Is it safe to use standard QFT during inflation? This question can be answered straightforwardly because both theories have been written in the same form. From equation (\ref{Teqn}) it is clear that the standard QFT is recovered in the regime in which the quantum aspects of the geometry can be neglected, and Section (\ref{sec:2.c}) provided the conditions under which this happens. When the background energy density $\langle \rho_o\rangle$  is below one thousandth of  $\rho_{P\ell}$, quantum corrections become negligible and General Relativity becomes an excellent approximation. This is the regime in which standard QFT arises from the more fundamental framework presented in this section. Therefore, in the inflationary era where  $\langle \rho_o\rangle  \lesssim 10^{-10} \rho_{P\ell}$, we expect the familiar QFT to be an excellent approximation.


By construction, this framework encompasses the Planck regime and is suitable to discuss trans-Planckian issues and distinguish real problems from apparent ones. In LQG there is a priori no impediment for trans-Planckian modes to exist. It may seem at first that the existence of a minimum area may preclude their existence, but quantum geometry is subtle and, for instance, there is no minimum value for volume or length. In addition, if we pay attention to the construction of the background quantum theory, trans-Planckian quantities appear there without causing problems: the value of the momentum $p_{(\phi)}$  of the background scalar field $\phi$ is generally large in Planck units. However, the background energy density is {\em bounded above} by a fraction of the Planck energy density. Something similar happens in our quantum field theory. There trans-Planckian modes are admitted {\em as long as the total energy density in perturbations remains small as compared to the background}. That is the real trans-Planckian problem, which becomes a non-trivial issue in the deep Planck regime where the volume of the universe acquires its minimum value.

\section{LQC extension of the inflationary scenario \label{sec:4}}

The previous sections have summarized the physical ideas and mathematical tools necessary to undertake the quantization of the sector of General Relativity containing the symmetries of cosmological space-times and the study of cosmic perturbation thereon. The goal of this section is to apply those techniques to extend the current picture of the evolution of our universe to include the Planck regime. 

The cosmological  $\Lambda$CDM model with an early phase of inflation contains conceptual limitations that are dictated by the domain of applicability of the physical theories in which it is based: General Relativity and Quantum Field Theory. One needs a theory of quantum gravity to extend the model to include physics at the Planck era. Subsection \ref{sec:4.a} summarizes how, by introducing a scalar field with suitable potential,  LQC provides a space-time in which the big-bang singularity is resolved by the quantum effects of gravity, and in which an inflationary phase arises almost unavoidably at later times. 
In subsection \ref{sec:4.a} it is shown how the evolution of cosmological perturbation can be extended to include the pre-inflationary space-times provided by LQC. In this sense the current scenario for the evolution of our universe and the genesis of cosmic inhomogeneities is extended all the way to the big bounce \cite{aan1}. This extension goes beyond the conceptual level, as it appears a narrow window in which the effects of Planck scale physics could be imprinted in the CMB and galaxy distributions, and concrete ideas  connecting those effects with forthcoming observations have been proposed.

\subsection{Inflation in LQC}
\label{sec:4.a}

As we have mentioned in previous sections, after the bounce there is a period of superinflation where $\dot{H}>0$ until the density reaches half its value at the bounce, after which
one has $\dot{H}<0$. It was first hoped that this period of superinflation
would be enough to account for the necessary number of {\it e-foldings} compatible with observations, but this period turns out to be too short when there is no potential for the scalar field. Thus, it is clear that one needs such a potential to compare the LQC predictions with the inflationary paradigm.
The simplest case one can consider is quadratic potential $\v(\phi)=(1/2)m^2\phi^2$, that has been extensively studied in the literature and is compatible with the 7-years WMAP observations \cite{wmap}. The existence of the bounce solves one of the conceptual challenges that the standard scenario, based on the GR dynamics poses. That is, in the GR dynamics, there is always a past singularity, even in the presence of eternal inflation \cite{bgv}. The standard formalism is therefore, conceptually incomplete. 

The question that we shall pose in this part is the following: Can we estimate how probable it is to have enough inflation for the cosmological background? Let us be more precise with the question. We know that every effective trajectory undergoes a bounce, and some of them will experience enough e-foldings and will be of phenomenological relevance. Rather amazingly, WMAP has provided us with a small observational window for the scalar field at the onset of inflation \cite{wmap,as3}, written in terms of a reference time $t_{k_*}$ for which 
a reference mode $k_*$ used by WMAP exited the Hubble radius in the early universe. With an $4.5\%$ accuracy, the data is, in Planck units \cite{wmap,as3}:
\[
\phi(t_{k_*})=\pm 3.15 \, ,\qquad \dot{\phi}(t_{k_*})=\mp 1.98 \times 10^{-7} \, ,
\qquad H(t_{k_*})=7.83\times 10^{-6} \, .
\]
We can now pose the question more precisely. From all the solutions $\mathbb{S}$ to the effective equations in
LQC, how many of them pass through the allowed interval? This poses yet another question. How are we going to `count' trajectories? Is there a canonical way of measuring them? A proposal to answer this question was put forward long ago \cite{ghs,hp} based on the idea of using the Liouville measure on phase space $\mathbb{S}$, that is invariant under time evolution.  The idea then is to compute the volume of $\mathbb{S}_{\rm wmap}$, those solutions that pass through the WMAP window, relative to the total volume of $\mathbb{S}$:
\be
{\rm Prob} =\f{{\rm Vol}(\mathbb{S}_{\rm wmap})}{{\rm Vol}(\mathbb{S})}\, .\label{prob-infla}
\ee
In order to  compute this probability, one has to be careful with the way one measures all
possible trajectories (for a discussion see \cite{sw}). 

Let us begin with the kinematical phase space $(V,\b;\phi,p_{(\phi)})$. The constrained surface $\bar{\Gamma}$ (as defined by the constraint ${\cal C}$) is three dimensional
and can be given coordinates $(V,\b,\phi)$. But in that surface, the symplectic structure is degenerate and does not define a volume form. For that one has to go to the space of physical states, or {\it reduced} phase space $\hat{\Gamma}$, formed by the gauge orbits on the constrained surface. An alternative is to perform a {\it gauge fixing} to select a two dimensional surface which is transversed only once by each gauge orbit. As we have seen
in the previous section the evolution of coordinate $\b$ is monotonous, so one can fix the gauge by
selecting $\b=\b_0$. With this choice, $\hat{\Gamma}$ has coordinates $(V,\phi)$. Now, the pullback
$\hat{\Omega}$ of $\Omega$ to $\hat{\Gamma}$ defines the Liouville measure there. The problem is that, with respect to this measure, the volume of $\hat{\Gamma}$ is infinite! One has to define a procedure
to `regularize' the integral to have finite results. The key observation is that, in the $k$=0 case we are considering, there is an extra gauge freedom that arises from the fact that the size of the
fiducial cell ${\cal V}$ one starts with is arbitrary. This means that a rescaling of the cell
${\cal V}\to \ell^3{\cal V}$ should leave the physics invariant. This rescaling translates into a rescaling of the canonical variables as $V\to \ell^3 V$, $p_{(\phi)}\to \ell^3 p_{(\phi)}$, while
$\b$ and $\phi$ remain invariant. However, this transformation on phase space does not leave the symplectic invariant, so it can not be regarded as a canonical transformation. Still, one has to
{\it gauge out} this symmetry in order to obtain truly physical quantities. The problem is that $\hat{\Omega}$ does not project down to the quotient. One possibility is to preform a further `gauge fixing' by selecting a cross section $\tilde{\mathbb{S}}$ of $\mathbb{S}$. For instance, one could choose a given value of volume $V=V_0$ (for our previous choice $\b=\b_0$). One can then restrict the measure to
the cross section $\tilde{\mathbb{S}}$ to obtain the measure $\d\tilde\mu$, that now depends only on $\phi$.
In effective LQC, one has
\be
\d\tilde{\mu}=\left[\f{3\pi}{\lambda^2}\,\sin^2(\lambda\b_0) - 8\pi^2\gamma^2\v(\phi)\right]^{1/2}\,\d\phi\, ,\label{measure-mu}
\ee
As expected, the GR measure is obtained by taking $\lambda\to 0$ \cite{ck-inflation}.
Even when the construction involved the Liouville measure that is invariant under Hamiltonian time evolution, the resulting measure $\d\tilde{\mu}$ on the space of physically distinguishable configurations $\tilde{\mathbb{S}}$ {\it depends on the choice of gauge fixing parameter} $\b_0$, in a non-trivial way \cite{ck-inflation,as3}.
 A choice of $\b$, in turn, fixes a value of the energy density
$\rho=\rho_0$, which implies that the probability will depend on the energy density at which it is
computed. In General Relativity there is no natural value of density for computing the probability, other than the big bang itself. The problem is that the density is infinite there and the range
of $\phi$ is unbounded, so the volume is also infinite. Another possibility would be to introduce a cut-off at, say, the Planck density \cite{klm}, but there is no reason to believe that GR is valid at that scale. In fact, one of the main lessons of loop quantum cosmology is that GR is not valid near the Planck scale (in energy density) but the isotropic degrees of freedom are rather described by the
effective LQC theory. In this description, there {\it is} a natural preferred density which is precisely the density at the bounce $\rcr$. Thus, in what follows we shall take the bounce as the natural point where to compute probability. The corresponding `gauge fixing' implementing this choice is then $\b_0=\pi/2\lambda$.

Let us now rephrase the question that we initially posed at the beginning of this part: What is the relative number of solutions $\tilde{\mathbb{S}}_{\rm wmap}$ that pass through the observational WMAP window, from the total number of solutions $\tilde{\mathbb{S}}$ at the bounce? As explained before, the probability is computed using formula (\ref{prob-infla}), where the volume is now obtained by integrating a uniform distribution (as a function of $\phi$). 
The key to computing the probability is then a detailed knowledge of the global dynamics, for all possible values $\phi_B$ of the scalar field
at the bounce. Extensive numerical evolutions have shown that almost all trajectories fall within
the observational window. It is only for the small window $-5.46< \phi_B < 0.934$ from the
total range of $\phi_B\in [-7.44 \times 10^{5},7.44 \times 10^{5}]$ that the future dynamics lies {\it outside} the WMAP window \cite{as3}. For this interval, the probability that the dynamics falls outside of the observational window is {\it less} than $3\times 10^{-6}$. To understand this, one
can see the LQC dynamics as shown in Fig.~\ref{Fig:1}, where one considers a uniform distribution at the bounce and follows the dynamics. As can be easily seen, most trajectories funnel into a very
small region that is precisely where the WMAP window is. Just before the onset of inflation the density is approximately $10^{-11}$ smaller than the density at the bounce. At that density the allowed WMAP region is only $4\%$ of the total allowed range in $\phi$ \cite{as3}. 
Thus, as seen in the Figure,
almost all of the trajectories starting at the bounce  scale fall into a very small region at the onset of inflation \cite{ck-inflation}. 

\begin{figure}[htb]
\centerline{\includegraphics [scale=0.35]{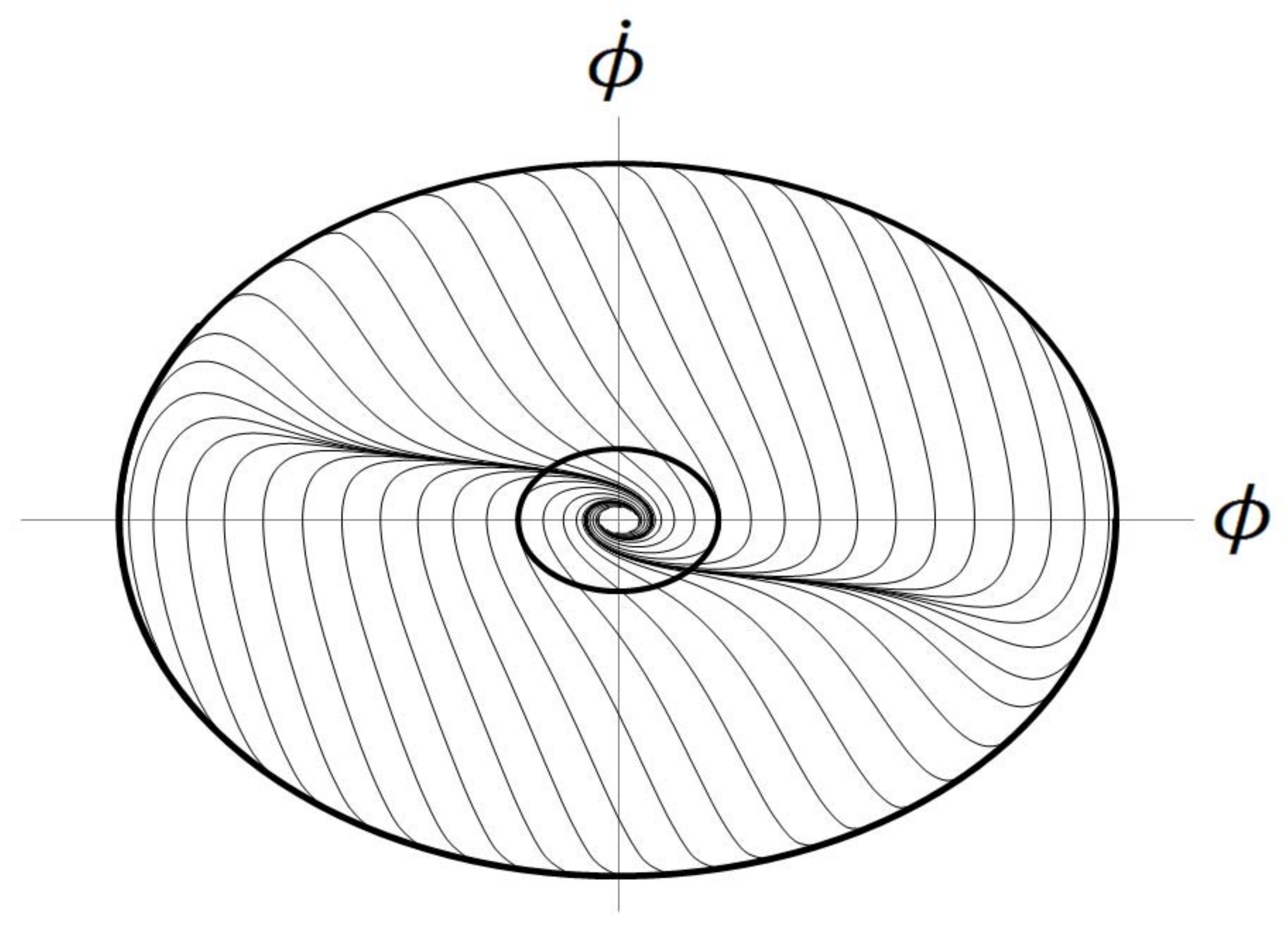}}
\caption{In this figure we plot the exterior, maximal density surface $\rcr$ and a surface of constant density $\rho_{\textrm{onset}}\ll \rho_{\textrm{max}}$ (not drawn to scale, of course) on the $(\dot\phi,\phi)$ plane. Trajectories with a uniform distribution at the LQC bounce ellipsoid  are plotted. 
Note that trajectories for which there is enough inflation get funnelled into a small region in the
smaller $\rho_{\textrm{onset}}$ ellipse. Near this surface, the GR and LQC dynamics almost coincide}
\label{Fig:1}
\end{figure} 

One should also note that this attractor feature of the global dynamics, 
together with the non-invariance of the measure $\d\tilde{\mu}$, 
explains why the probability is much smaller when computed in General Relativity at the onset of inflation \cite{gt,ck-inflation}.

Let us summarize. In LQC it is natural to consider the bounce as the point where to compute probability of inflation. The global dynamics is such that most of the trajectories get funnelled
into the small WMAP window at the onset of inflation where the density is 11 orders of magnitude smaller than the density at the bounce. Thus, one can conclude that having enough inflation is generic in loop quantum cosmology for the homogeneous and isotropic background,
when semiclassical states are considered.

\subsection{Pre-inflationary evolution of cosmic perturbations}
\label{sec:4.b}

In this section we apply the quantum theory of cosmological perturbations on the quantum, pre-inflationary space-time to extend the study of cosmic inhomogeneities all the way back to the Planck era. 
In addition to the {\em conceptual} completion provided by the inclusion of Planck scale physics, the resulting framework  opens an exciting avenue to extend observations into the Planck regime. Before entering into technical details, we summarize here the physical idea behind this possibility. 

It is known since the seminal work by Parker in the 60's \cite{parker66,parker69}, that a dynamical expansion of the universe is able to excite quanta, or `particles', of  test fields out from an initially vacuum state. This phenomenon of particle creation is one of the main features of QFT in curved space-times, and plays a key role in black hole thermal radiance and in the generation of cosmic inhomogeneities during inflation. If $\vec{k}$ represents a co-moving Fourier mode of a test scalar field in FLRW, excitations on that mode may be created if the energy scale provided by the space-time scalar curvature is comparable to the physical wavelength  $\lambda = 2\pi a/k$ at some time during the evolution. The amount of quanta created during a period of expansion in each mode depends on the details of the scale factor $a(t)$ as a function of time. Let us focus on the finite range of momenta that is accessible in cosmological observations. The previous argument tell us that, even if those modes are `born' in the ground state at  time of the bounce, particles may be created during the evolution. The resulting state, e.g. at the onset of inflation, would then depart from the vacuum state at that time as a consequence of the non-trivial evolution, and the spectrum of particles created will carry information about the pre-inflationary space-time geometry. 

Furthermore, it has been shown in the context of inflation that {\em the predictions for the CMB and the distribution of galaxies are sensitive to the details of the state describing perturbations at the onset of inflation} \cite{holman-tolley,agullo-parker,ganc,agullo-navarro-salas-parker}, and concrete observation have been proposed that could reveal information about that state \cite{halo-bias1,halo-bias2,halo-bias3}. In other words, those observations may reveal information about the propagation of perturbations {\em before} inflation, when quantum gravity corrections dominate. 

In the inflationary scenario observable modes have wavelength much smaller than the radius of curvature at the onset of inflation (in the cosmological argot, modes are deeply inside the Hubble radius). The sometimes implicit assumption in inflationary physics is that, whatever happened before inflation, wavelength of interest were much smaller than the radius of curvature {\em at any time before inflation}. Under this assumption, pre-inflationary dynamics  for those modes is indistinguishable from  an evolution in Minkowski space-time, and the use of a vacuum state is justified.  The relevant question is then: is this assumption accurate in the pre-inflationary background provided by LQC? More explicitly, consider modes with physical wavelength smaller that the radius of curvature at the beginning of inflation, and propagate them backward in time until the bounce. Do those wavelength generically remain smaller that the radius of curvature of the dressed metric $\tilde g_{ab}$  during the entire pre-inflationary evolution? The detailed analysis of \cite{aan1,aan3} shows that the answer to this question is in the negative (see Fig.~\ref{Fig:2}). 
While short enough wavelengths (large enough momenta) remain always smaller that the curvature radius, there are modes which at some time during the evolution have physical size comparable to it. The evolution of those modes {\em is} sensitive to the space-time curvature and the quantum state at the onset of inflation will depart from the vacuum.

\begin{figure}[htb]
\centerline{\includegraphics[width=10cm]{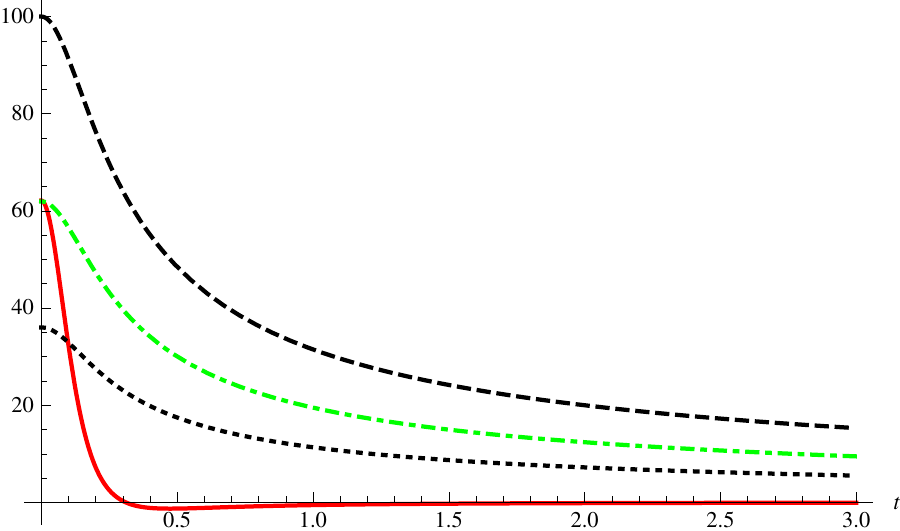}}
\caption{This plot shows: i) The scalar curvature of the effective geometry (red solid line), ii)  The physical momentum squared $(k/\tilde{a}(t))^2$, for $k=6$ (dotted black line), and $k=10$  (dashed black line), and iii)  $(k_R/\tilde{a}(t))^2$, where $k_R$ is the co-moving scale associated with the maximum value of the curvature (dotted-dashed green line); as a function of cosmic time $t$. By convention, we choose the scale factor of the effective geometry to be one at the bounce, $\tilde a(0)=1$. Both axes are in Planck units. Curvature attains the maximum value at the bounce and decreases very fast after it. Modes with momentum $k$ larger than the scale of curvature at the bounce, $k > k_R$, have physical momentum larger than the curvature during the entire evolution (dashed black line). Those modes do not `feel' the curvature and evolve as if they were in Minkowski space-time. On the other hand, modes that at the bounce have physical momentum smaller that the curvature, $k<k_R$, quickly evolve to become of the same order as the curvature scale (black dotted line), and therefore their evolution will differ considerably from that in flat space. At later times those modes also become two energetic to feel the space-time curvature.}
\label{Fig:2}
\end{figure}

Notice that in LQC the maximum value of the curvature takes place at the bounce time and this value is universal, fixed by the quantum geometry and independent of the form of the scalar field potential. If we call $k_R$ the co-moving scale associated with this maximum value of the curvature, we expect  excitations with $k\lesssim k_R$ to be created during the evolution, concretely in the Planck regime near the bounce.  On the other hand, for modes with $k\gg k_R$ pre-inflationary dynamics has negligible effect. From this qualitative discussion we may expect observable effects from Planck scale physics in CMB and large scale structure if observations are accessible to modes $k$ around or smaller than the universal scale $k_R$ provided by LQC. 

In the remainder of this section we provide precise computations that support this qualitative physical picture. We start by specifying the initial condition for both background and perturbations at the bounce. We then evolve those perturbations until the end of slow-roll inflation, compute the resulting quantum state and the power spectrum for scalar and tensor perturbations, and study under what set of initial conditions quantum gravity corrections may be sizeable for observable modes.

\subsubsection{Initial Conditions}
\label{sec:4.b.1}

In the standard inflationary paradigm one specifies `initial data' for the background and perturbations at the onset of slow-roll. From a fundamental point of view, it would be more satisfactory to impose initial conditions at the `beginning' rather than at an intermediate time in the evolution of the universe. In classical cosmology the `beginning' is the big bang singularity, and it is not possible to unambiguously defined initial condition at that time. In LQC the big bang is replaced by a quantum bounce where physical quantities do not blow up, providing a preferred time to specify initial data.

In the test field approximation, the total wave function naturally decomposes as a product $\Psi=\Psi_0\otimes \psi$, and this form holds as long as back-reaction of perturbations remains negligible. We need therefore to specify initial data for both, $\Psi_0$ and  $\psi$.\\

$\bullet$ {\em Background.} 
 For computational purposes, it is convenient to make the following further simplification on the background dynamics. As described in section~\ref{sec:2.a}, the background wave function $\Psi_0$ can be chosen to be highly peaked along  the entire evolution, including the deep Planck regime. The `peak' of that wave function describes an effective geometry characterized by the scale factor $\bar a(\phi)=\langle \h a(\phi)\rangle$, which satisfies the effective equation (\ref{eff-fried}). Because the dispersion of $\Psi_0$ remains very small during evolution, it is convenient to ignore quantum fluctuations in our computations, by making a `mean field' approximation in which the expectation values of powers of background operators, such as $\h a$ and $\h H_o$, are replaced by the same powers of their expectation. For instance, in the evolution of quantum inhomogeneities given by Eq. (\ref{Teqn}), this is equivalent to replace $\tilde a \approx \bar a$. At the practical level this is an excellent approximation, e.g. numerical errors in simulations turn out to be larger  than those introduced by the mean field approximation. 

In subsection \ref{sec:4.a} we described the effective pre-inflationary  background arising in LQC for the representative example of a quadratic potential. In that effective geometry  initial data is entirely specified by the value of the scalar field at the bounce, $\phi_B$, and, unless $\phi_B$ lies in a small region $R$ around $\phi_B$=0, the evolution generically finds an inflationary phase at late times compatible with WMAP observations \cite{wmap}. Therefore, we will choose $\Psi_0$ to be a state sharply peaked in an effective trajectory specified by a value of $\phi_B$ that lies outside the region $R$. 
 
The effect of choosing different values of $\phi_B$ can be understood using the effective equations (\ref{eff-fried}) together with numerical simulations. On the one hand, immediately after the bounce the background evolution is entirely dominated by quantum gravity effects, and it is largely insensitive to the concrete value of $\phi_B$. Except for very small momenta $k$, the times at which perturbations $\Q_k$ and $\T_k$ `feel' the space-time curvature is precisely just  after the bounce (see Fig.~\ref{Fig:2}). Therefore, the features that those modes acquire during the evolution turn out to be quite insensitive to the value of $\phi_B$. On the other hand, different values of $\phi_B$ do modify significantly the space-time geometry at later times. The larger $\phi_B$, the longer it takes to reach the end of slow-roll inflation, or, equivalently, the larger the amount of expansion of the universe between the bounce and the end of slow-roll. A larger amount of expansion implies that observable modes had larger physical momentum at the time of the bounce. Because by convention {\em we fix the scale factor at the bounce} $\bar a_B=1$ (rather than $\bar a_{\rm today}=1$), the effect of choosing different values of $\phi_B$  essentially translates into a change in the range of co-moving momenta $k$ relevant for observations, moving to larger $k$'s as $\phi_B$ increases. If $[k_{\rm{min}}$, $k_{\rm{max}}\approx 2000 k_{\rm{min}}]$ is the window covered by WMAP, we have, for instance, $k_{\rm{min}}\approx 2.8\times 10^{-3}$ for $\phi_B=1$, $k_{\rm{min}}\approx 0.14$ for $\phi_B=1.1$ and  $k_{\rm{min}}\approx 8.2$ for $\phi_B=1.2$. The physical momentum  $k/\bar a_{\rm today}$ of modes observed today is of course the same in all cases, but the convention $\bar a_B=1$ makes that different amount of expansion (i.e. different $\phi_B$)  translates  into different co-moving $k$ for those modes.\\

$\bullet$ {\em Perturbations.} As already occurs in classical space-times, quantum fields in quantum cosmological backgrounds does not admit a preferred state that we can call {\em the vacuum}. In backgrounds with large enough number of isometries, e.g. Minkowski or de Sitter space-time, a preferred ground state can be singled out by imposing symmetry in combination with regularity conditions. In our quantum FLRW  we follow the same criteria, and look for quantum states $\psi$ invariant under the  isometries of the background, spatial translations and rotations, with appropriate ultraviolet behavior. In section \ref{hilbertspace} we summarized the construction of the Hilbert space $\H_1$ of 4th-order adiabatic states. In $\H_1$, the family of 4th-order adiabatic vacua is the preferred set of initial conditions selected by symmetry and regularity requirements. This is the set of initial data we choose for perturbations. As opposed to Poincare or de Sitter invariance, symmetry under spatial translations and rotations is not restrictive enough to select a unique state, but it substantially narrows down the possibilities. The next subsection will summarize the time evolution of different choices of initial state within the family of 4th-order adiabatic vacua, and will show that quantities of interest such as the power spectrum of observable modes, are all very similar.
%

Physically, the choice of a 4th-order adiabatic vacuum at the time of the bounce corresponds to assume `initial quantum homogeneity'. One is requiring that the portion of the universe corresponding to our observable patch at the time of the bounce is {\em as homogeneous as quantum mechanics allows}, i.e. only vacuum fluctuation of inhomogeneities are present. This is a strong assumption. The motivation comes from \cite{aan1,aan3}:

\begin{itemize}
\item In a universe containing a phase of inflation lasting at least for 60 $e$-folds,  the physical size of observable universe was very small at the bounce time, $\lesssim  10 \ell_{\rm Pl}$,  for the solutions of interest.
\item The `quantum degeneracy force' responsible of the bounce has a diluting effect that may produce homogeneity at scales of the order of the Planck length at the bounce. This is the new ingredient that LQC provides at the time of the bounce to produce homogeneity at Planck scale distances.
\item There is a precise sense in which the assumption of quantum homogeneity captures a quantum version of the Weyl curvature hypothesis \cite{penrose-weyl}.

\end{itemize}

\subsubsection{Power Spectrum}
\label{sec:4.b.2}

Our task is to use the equations of the quantum theory summarized in section \ref{QFTQST} to compute the state of cosmic inhomogeneities at the end of the inflationary epoch, by starting from the initial condition specified above for background and perturbations at the time of the bounce. 

Due to computational limitations, it is convenient to restrict numerical simulations to backgrounds  for which the bounce is kinetic energy dominated, where it has been shown that quantum fluctuations of $\Psi_0$ remain very small along the entire evolution. Several numerical simulations have been carried out for effective backgrounds with initial conditions $\phi_B\in(0.93,1.5)$, which turns out to be the most interesting range \cite{aan3}. It is not expected that new features appear for larger values of $\phi_B$, but computational limitations make difficult to check it explicitly. 

For perturbations, simulations have been carried out using different choices of 4th-order adiabatic vacua, and the results are all very similar. Fig.~\ref{Fig:3} and \ref{Fig:4}  are obtained by using the `obvious' or `standard' 4th-order vacuum at the bounce time $\tilde\eta_B$ (see \cite{aan2} for precise definition), and they show the relevant information of the evolved state. 

\begin{figure}[htb]
\centerline{\includegraphics[width=11cm]{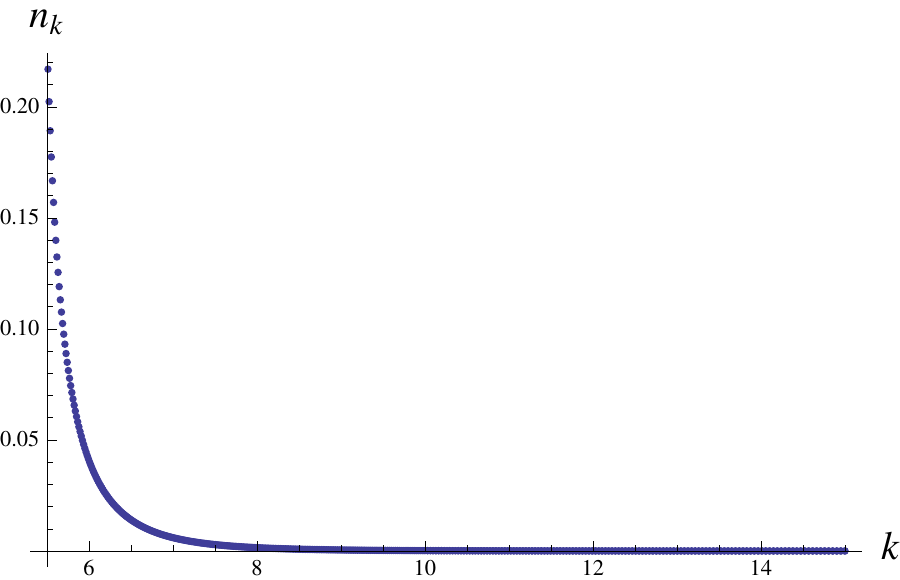}}
\caption{ Number $n_k$ of scalar  `excitations/particles' with  comoving momentum $\vk$ in the interval $[\vk,\vk+d\vk]$, per comoving unit volume contained in the evolved state as compared to the BD vacuum during inflation. The plot is computed for $\phi_B=1.15$ and for the `obvious' 4th-order adiabatic vacuum at the bounce. The horizontal axes is in Planck units.}
\label{Fig:3}
\end{figure}

\begin{figure}[htb]
\centerline{\includegraphics[width=10cm]{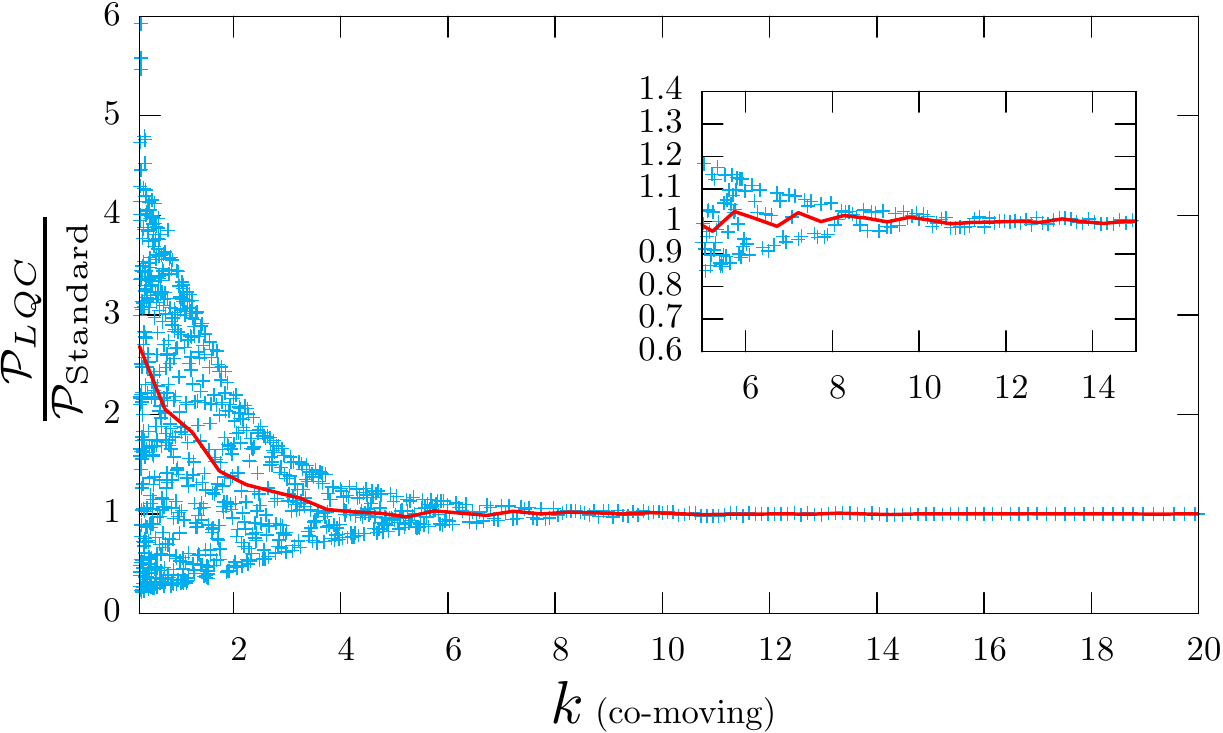}}
\caption{Ratio of the LQC power spectrum for scalar perturbation to the standard inflationary power spectrum. Crosses show the ratio for different values of $k$.  The LQC power spectrum oscillates rapidly for small $k$. The solid curve averages over bins of width $\Delta k=0.5$. The inset shows a zoom-in of the interesting region around $k$=9.}
\label{Fig:4}
\end{figure}

First of all, to gain intuition on the effect of the pre-inflationary evolution, we compare the evolved state with the natural vacuum during inflation, the so-called Bunch-Davies (BD) vacuum.  
Fig.~\ref{Fig:3} shows the number $n_k$ of `excitations/particles' with momentum $\vk$ per comoving unit volume in space and momentum,  contained in the evolved state relative to the BD vacuum during inflation. The plot is computed for $\phi_B=1.15$ but, as explained in subsection \ref{sec:4.b.1}, it is not altered by choosing a different value inside our family. Changing the value of $\phi_B$ has essentially the effect of shifting the location  of the observationally relevant window $[k_{\rm min}, k_{\rm max}\approx 2000 k_{\rm min}]$ in the horizontal axes of the plot, which moves steadily to the right as $\phi_B$ increases. Fig.~\ref{Fig:3} is in good agreement with the qualitative arguments presented at the beginning of section \ref{sec:4.b}. Namely, the pre-inflationary evolution affects modes with low $k$, for which a considerable amount of excitations have been `created'. On the contrary,  modes with  large $k$ remain in the ground state at the onset of inflation. As it was expected, for $k> k_R\approx 7.7$ (recall that $k_R$  is the comoving scale associated with the scalar curvature of the effective metric at the bounce), the number of BD particles contained in the evolved state is very close to zero. Therefore, if $k_{\rm min} \gtrsim k_R$,  that corresponds to $\phi_B\gtrsim 1.2$, the evolved state is indistinguishable from the BD vacuum for observable modes. For  $\phi_B\lesssim 1.2$ the state at the onset of inflation differs significantly from the vacuum for modes in the interesting window and, as analyzed in detail in \cite{holman-tolley,agullo-parker,ganc,agullo-navarro-salas-parker}, those deviations have an important effect on the predictions of inflation for the spectrum of cosmic inhomogeneities, specially regarding non-Gaussianity. There exist concrete proposals for observables in the CMB \cite{halo-bias1,halo-bias3} and in the distribution of galaxies \cite{halo-bias1,halo-bias2} that should be sensitive to the effects of the created particles. 

A quantity of direct observational interest is the power spectrum of tensor and scalar perturbations
\be  P_{\T}(k)=\hbar \frac{k^3}{2\pi^2} |e_k|^2 \, , \quad \quad P_s(k)=\hbar \frac{k^3}{2\pi^2} \left(\frac{\dot\phi}{H}\right)^2 |q_k|^2 \, , \ee
where all quantities are evaluated at the end of inflation, $H$ is the Hubble rate, and $q_k(t)$ and $e_k(t)$ are the Fourier modes of scalar and tensor perturbations, respectively.
Fig.~\ref{Fig:4}  shows the relation between the LQC power spectrum computed with the evolved state and the standard inflationary power spectrum that assumes the BD-vacuum, for scalar perturbations. The conclusions are  similar to the ones obtained from Fig.~\ref{Fig:3}, namely for $\phi_B\gtrsim 1.2$ the power spectrum of observable modes is indistinguishable from the standard inflationary predictions. For smaller values of $\phi_B$ deviations become sizable for modes of observational interest. For instance, for $\phi_B=1.15$ we have $k_{\rm min}\approx 1$ and deviations from standard prediction will appear for modes with $\ell\lesssim 30$ in the WMAP angular decomposition. These deviations are inside current uncertainties. However, the fact that the state for perturbations differs from the BD-vacuum opens a window to observe those effects. 

The analogous plot for tensor modes has the same form as Fig.~\ref{Fig:4}, and the conclusions are also the same \cite{aan3}. In particular, there are no important corrections for the tensor-to-scalar ratio, although the inflationary consistency relation, which relates the tensor-to-scalar-ratio and the tensor spectral index, is modified \cite{aan3}.

\subsubsection{Self-consistency}

The last step is to check whether there exist a big enough set of physical states $\psi$ on the Hilbert space for which the truncation underlying our quantum theory, the test field approximation, holds during the entire evolution. This is an intricate question because: i) It requires a detailed analytical control of the necessary regularization on states and composite operators on our Hilbert space; ii) Numerical implementation of those techniques are necessary to check self-consistency {\em at any time during the evolution}, dealing with the subtleties of having numerical control on the subtraction of quantities that tend rapidly to infinity, during a period that covers around $11$ orders of magnitude in energy density.  

Section  \ref{QFTQST} summarized the necessary tools to check self-consistency and pointed out that a sufficient condition is that the energy density in perturbations $\langle \h \rho \rangle$ be negligible compared to the background $\langle \h \rho_o \rangle$ {\em at any time} during the evolution. Fig.~\ref{Fig:5} shows the result of the numerical evolution of the energy density for scalar perturbations (analogous results hold for tensor perturbations). The plot shows the ratio $\langle \h \rho_{\Q} \rangle/\langle \h \rho_o \rangle$ for a background corresponding to $\phi=1.23 $ and the `obvious' 4th adiabatic order vacuum specified at the bounce. This ratio remains small for the entire evolution, including the Planck regime. The initial condition $\phi=1.23$ corresponds to $k_{\rm min}\approx 30$, therefore the number of excitations over the BD state on observable modes is negligible (see Fig.~\ref{Fig:3}) for this background.
Additionally, there exist an analytical argument \cite{aan3} ensuring that, given a state for perturbations for which back-reaction is negligible, there exist a well defined neighborhood of that state with the same property. Each of those provide a state at the beginning of slow-roll indistinguishable from the BD vacuum. They provide therefore, viable extensions of the standard inflationary scenario that includes Planck scale physics \cite{aan1,aan3}. 

\begin{figure}[htb]
\centerline{\includegraphics[width=10cm]{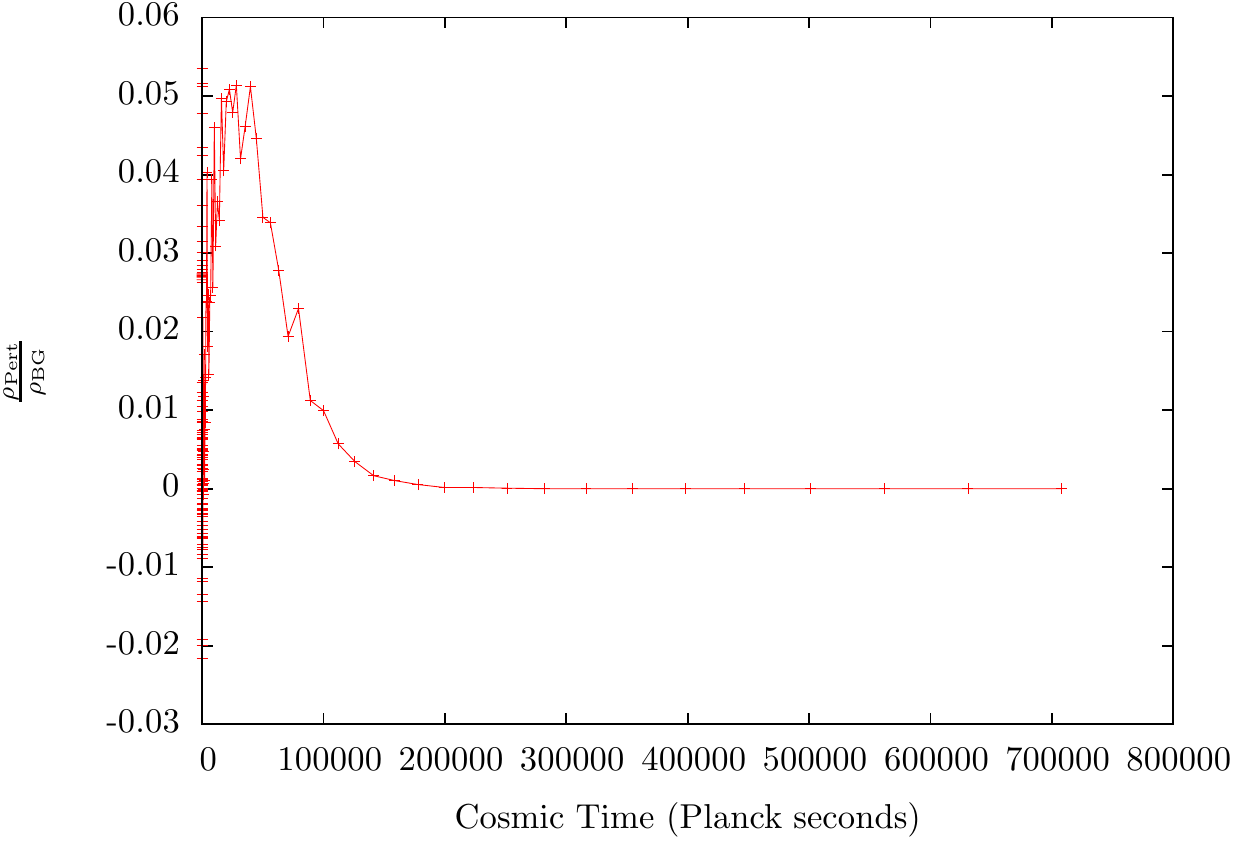}}
\caption{Ratio of the energy density of scalar perturbation to the background energy density as a function of cosmic time. The initial conditions were chosen as $\phi_B=1.23$ for the background, and the `obvious' 4th-order adiabatic vacuum at the bounce for perturbations. Slow-roll inflation starts about $3\times 10^5$ Planck seconds after the bounce. During the entire evolution the ratio remains small. This example constitutes a self-consistent extension of the evolution of cosmic inhomogeneities to include the Planck era.}
\label{Fig:5}
\end{figure}

For the range $\phi_B<1.2$ there are only upper bounds for $\langle \h \rho_{\Q} \rangle$ which are far from being optimal. At the present time there are no explicit computations for which the test field approximation is satisfied for $\phi_B$ in that window, and additional work is required to establish the self-consistency of our truncation scheme.

\section{Conclusions}

One of the most pressing questions a quantum theory of gravity has pertains to both theoretical and observational issues in cosmology. In the theoretical front the standard model is based on General Relativity that possesses an initial singularity, a signal that the theory breaks down at some point. On the observational front, the CMB spectrum poses very stringent conditions for any theory of the early universe. One of such scenarios is given by the inflationary paradigm, that explains very successfully the detailed structure of the inhomogeneities seen in the CMB as an imprint of quantum fluctuations of certain fields
just before the inflationary phase. Can one have a formalism that provides a satisfactory, nonsingular description both at the Planck scale and at the onset of inflation? Interestingly, loop quantum cosmology allows to answer both questions in the affirmative.

As we have described in this Chapter, when one considers the homogeneous degrees of freedom, the so called `background geometry', the formalism provides precise singularity resolution, replacing the classical big bang with a big bounce. The dynamics of semiclassical states is very well described by an effective theory that captures the leading quantum gravity effects and allows one to describe the spacetime geometry in terms of an effective background metric.

The inflationary scenario is very powerful to explain in great detail many features of the observed CMB spectrum. It is however, incomplete in various directions. In particular, it is based on General Relativity where the spacetimes under consideration are past incomplete, that is, singular.
As we have described in detail, one can indeed extend the scenario back in time to the Planck scale. For that one needs two new ingredients. The first one is a formalism that allows one
to treat quantum perturbations of the spacetime metric propagating not on a classical spacetime, but rather on a {\em  quantum} spacetime. 
The second ingredient involves consistency conditions that ensure us that one can `evolve' the quantum perturbations back to the Planck scale without violating the approximations that yield validity to the formalism. As we have seen one can indeed consistently consider the 
extension of the inflationary scenario.

Perhaps the most important question is whether this extension to the quantum bounce provides a window for Planck scale physics to be observed in the CMB. As we have described, the sector of the parameter space that has been explored provides predictions that are fully consistent with the standard inflationary scenario, under current observations. Further explorations are needed to decide whether the scenario provided by LQC is both consistent in the full parameter space and provides us with distinct testable predictions.

\section*{Acknowledgements}

We would like to thank A. Ashtekar, P. Singh and W. Nelson for discussions and collaboration. I.A. thanks the Marie Curie program of the EU for funding. This work was partly funded by DGAPA-UNAM IN103610, CONACyT CB0177840, and NSF PHY0854743 grants and by the Eberly Research Funds of Penn State.


\begin{thebibliography}{99}

\bibitem{lqg} A.~Ashtekar and J.~Lewandowski, {Background
    independent quantum gravity: A status report}, Class. Quant.
    Grav. {\bf 21} R53-R152 (2004);
    T.~Thiemann, {\em Introduction to Modern
    Canonical Quantum General Relativity.} (Cambridge University Press,
    Cambridge, (2007))
    
    \bibitem{abl} A.~Ashtekar, M.~Bojowald and J.~Lewandowski,
    {Mathematical structure of loop quantum cosmology}. Adv. Theo.
    Math. Phys. \textbf{7} 233--268 (2003)

\bibitem{akl} A.~Ashtekar, W.~Kaminski and J.~Lewandowski, Quantum
    field theory on a cosmological, quantum space-time, Physical
    Review D\textbf{79} 064030 (2009), arXiv:0901.0933.


\bibitem{aan2}I.~Agullo, A.~Ashtekar and W.~Nelson, An extension of the quantum theory of cosmological perturbations to the Planck era, Phys. Rev. D\textbf{87},  043507 (2013), \text{arXiv:1211.1354}.


\bibitem{bgv} A.~Borde, A.~Guth and A.~Vilenkin, Inflationary
    space-times are not past-complete, Phys. Rev. Lett.
    \textbf{90} 151301 (2003).
    
 
 \bibitem{ach} A.~Ashtekar, M.~Campiglia and A.~Henderson,
  Casting Loop Quantum Cosmology in the Spin Foam Paradigm,
  Class.\ Quant.\ Grav.\  {\bf 27}, 135020 (2010);
  Path Integrals and the WKB approximation in Loop Quantum Cosmology,
  Phys.\ Rev.\ D {\bf 82}, 124043 (2010).

\bibitem{vidotto} C.~Rovelli and F.~Vidotto,
  On the spinfoam expansion in cosmology,
  Class.\ Quant.\ Grav.\  {\bf 27}, 145005 (2010);
  E.~Bianchi, C.~Rovelli and F.~Vidotto,
  Towards Spinfoam Cosmology,
  Phys.\ Rev.\ D {\bf 82}, 084035 (2010).
  
   \bibitem{hybrid1} M.~Martin-Benito, L.~J.~Garay and G.~A.~Mena
    Marugan, Hybrid quantum Gowdy cosmology: combining loop and
Fock quantizations, Phys. Rev. D\textbf{78} 083516 (2008);\\
L.~J.~Garay, M.~Martn-Benito, G.~A.~Mena Marugan, Inhomogeneous loop
quantum cosmology: Hybrid quantization of the Gowdy model, Phys.
Rev. D\textbf{82} 044048 (2010).
  
\bibitem{hybrid2} D.~Brizuela, G.~A.~Mena Marugan and T.~Pawlowski,
    Big bounce and inhomogeneities, Class. Quant. Grav. \textbf{27}
    052001 (2010).

\bibitem{hybrid3} M.~Martin-Benito, G.~A.~Mena Marugan,
    E.~Wilson-Ewing, Hybrid quantization: From Bianchi I to the
    Gowdy model, Phys. Rev. D\textbf{82} 084012 (2010).

\bibitem{hybrid4} M.~Martin-Benito, D.~Martin-de Blas,
    G.~A.~Mena Marugan, Matter in inhomogeneous loop quantum cosmology: the
    Gowdy $T^3$ model, \texttt{arXiv:1012.2324}.

\bibitem{hybrid5} D.~Brizuela, G.~A.~Mena Marugan and T.~Pawlowski,
    Effective dynamics of the hybrid quantization of the Gowdy T3
    universe, \texttt{arXiv:1106.3793}. 

  
     
 \bibitem{brizuela} D.~Brizuela, D.~Cartin and G.~Khanna,
  Numerical techniques in loop quantum cosmology,
  SIGMA {\bf 8}, 001 (2012).
    
 \bibitem{pert_tensor1} M.~Bojowald and G.~M.~Hossain, Loop quantum
    gravity corrections to gravitational wave dispersion, Phys. Rev.
    D {\bf 77} 023508 (2008).
 
\bibitem{ns_inflation} W.~Nelson and M.~Sakellariadou,
  Lattice refining loop quantum cosmology and inflation,
  Phys.\ Rev.\  D {\bf 76} 044015 (2007).
  
  
\bibitem{barrau1} J.~Grain and A.~Barrau, Cosmological footprints of
    loop quantum gravity, Phys. Rev. Lett. \textbf{102} 081301
    (2009).
        
\bibitem{barrau2} J.~Grain, T.~Cailleteau, A.~Barrau and A.~Gorecki,
    Fully loop-quantum-cosmology-corrected propagation of gravitational
    waves during slow-roll inflation, Phys. Rev. D\textbf{81} 024040
    (2010).

\bibitem{barrau3} J.~Mielczarek, T.~Cailleteau, J.~Grain and
    A.~Barrau, Inflation in loop quantum cosmology: Dynamics and spectrum of
gravitational waves, Phys. Rev. D\textbf{81} 104049 (2010).

\bibitem{barrau4} J.~Grain, A.~Barrau, T.~Cailleteau and
    J.~Mielczarek, Observing the big bounce with tensor modes in the
    cosmic microwave background: Phenomenology and fundamental LQC
    parameters, Phys. Rev. D\textbf{82} 123520 (2010).
    
   \bibitem{bojowald&calcagni} M. Bojowald, G. Calcagni and S. Tsujikawa,  Observational test of inflation in loop quantum cosmology, JCAP 1111 (2011) 046. 


\bibitem{barrau5} T.~Cailleteau, J.~Mielczarek, A.~Barrau and
    J.~Grain, Anomaly-free scalar perturbations with holonomy corrections in
    loop quantum cosmology, Class.Quant.Grav. 29 (2012) 095010.
 
 \bibitem{madrid} M.~Fernandez-Mendez, G.~A.~Mena Marugan, and J.~Olmedo, Hybrid quantization of an inflationary universe, Phys. Rev. D. {\bf 86},  024003 (2012).
    
\bibitem{wilson-ewin} E.~Wilson-Ewing, Lattice loop quantum cosmology: scalar perturbations, Class.Quant.Grav. 29 215013 (2012); The Matter Bounce Scenario in Loop Quantum Cosmology,
  \texttt{arXiv:1211.6269}.
    
    
\bibitem{asrev} A.~Ashtekar and P.~Singh, Loop quantum cosmology: A
    status report, Class. Quant. Grav. 28, 213001 (2011).
    
\bibitem{lqcreview} K.~Banerjee, G.~Calcagni and M.~Martin-Benito,
  Introduction to loop quantum cosmology,
  SIGMA {\bf 8}, 016 (2012)
  \texttt{arXiv:1109.6801}
  
\bibitem{singh-numerical} P.~Singh,
  Numerical loop quantum cosmology: an overview,
  Class.\ Quant.\ Grav.\  {\bf 29}, 244002 (2012).
  
   \bibitem{calcagni} G. Calcagni, Observational effects from quantum cosmology, \texttt{arXiv:1209.0473}.



\bibitem{aps3} A.~Ashtekar, T.~Pawlowski and P.~Singh, {Quantum
    nature of the big bang: Improved dynamics}, Phys. Rev. D{\bf
    74} 084003 (2006).

\bibitem{acs}A.~Ashtekar, A.~Corichi and P.~Singh, {Robustness
    of predictions of loop quantum cosmology}, Phys. Rev.
    D\textbf{77} 024046 (2008). 
  
  \bibitem{afw} A.~Ashtekar, S.~Fairhurst and J.~Willis, {Quantum
    gravity, shadow states, and quantum mechanics}, Class.\ Quantum\
    Grav. \textbf{20} 1031-1062 (2003).
  
  


\bibitem{ach4}  A.~Ashtekar and M.~Campiglia,
  ``On the Uniqueness of Kinematics of Loop Quantum Cosmology,''
  Class.\ Quant.\ Grav.\  {\bf 29}, 242001 (2012)
  \texttt{arXiv:1209.4374}.

\bibitem{lost} J.~Lewandowski, A.~Okolow, H.~Sahlmann and
    T.~Thiemann, {Uniqueness of diffeomorphism invariant states on
    holonomy flux algebras}, Comm. Math. Phys. \textbf{267} 703-733
    (2006).

\bibitem{cf} C.~Fleishchack, {Representations of the Weyl algebra in
    quantum geometry}, Commun. Math. Phys. \textbf{285} 67-140
    (2009).
  
  \bibitem{aps2} A.~Ashtekar, T.~Pawlowski and P.~Singh, {Quantum
    nature of the big bang: An analytical and numerical
    investigation}, Phys. Rev. {\bf D73} 124038 (2006).
    
    
\bibitem{cs2} A.~Corichi and P.~Singh, Quantum bounce and cosmic
    recall, Phys. Rev. Lett. {\bf 100} 209002 (2008)

\bibitem{kp2}W.~Kaminski and T.~Pawlowski, Cosmic recall and the
    scattering picture of loop quantum cosmology, Phys. Rev.
    D\textbf{81} 084027 (2010)

\bibitem{cm1} A.~Corichi and E.~Montoya,
  On the Semiclassical Limit of Loop Quantum Cosmology,
  Int.\ J.\ Mod.\ Phys.\ D {\bf 21}, 1250076 (2012)
  \texttt{arXiv:1105.2804};
  Coherent semiclassical states for loop quantum cosmology,
  Phys.\ Rev.\ D {\bf 84}, 044021 (2011)
  \texttt{arXiv:1105.5081}.
    
    
\bibitem{apsv} A.~Ashtekar, T.~Pawlowski, P.~Singh and
    K.~Vandersloot, {Loop quantum cosmology of k=1 FRW
    models}. Phys. Rev. D\textbf{75} 0240035 (2006).

\bibitem{warsaw1}L.~Szulc, W.~Kaminski, J.~Lewandowski, Closed
    FRW model in loop quantum cosmology, Class.\ Quant.\ Grav.\ \textbf{24} 2621
   (2007).    
    
    
\bibitem{ck2} A.~Corichi and A.~Karami,
  Loop quantum cosmology of k=1 FRW: A tale of two bounces,
  Phys.\ Rev.\ D {\bf 84}, 044003 (2011)
  \texttt{arXiv:1105.3724}.

\bibitem{k=-1}  K.~Vandersloot,
  Loop quantum cosmology and the k = - 1 RW model,
  Phys.\ Rev.\ D {\bf 75}, 023523 (2007)
  [gr-qc/0612070].

\bibitem{szulc} L.~Szulc,
  Open FRW model in Loop Quantum Cosmology,
  Class.\ Quant.\ Grav.\  {\bf 24}, 6191 (2007)
  [arXiv:0707.1816 [gr-qc]].
    
    
\bibitem{bp} E.~Bentivegna and T.~Pawlowski, Anti-deSitter
    universe dynamics in LQC,  Phys. Rev. D\textbf{77} 124025
    (2008).
    

\bibitem{kp1}W.~Kaminski and T.~Pawlowski, The LQC evolution
    operator of FRW universe with positive cosmological constant
    Phys. Rev. D\textbf{81} 024014 (2010).

\bibitem{ap} A.~Ashtekar and T.~Pawlowski, Loop quantum
    cosmology with a positive cosmological constant, Phys. Rev.
    \text bf{85}, 064001 (2012).
    
    
\bibitem{we} E.~Wilson-Ewing, Loop quantum cosmology of Bianchi type
    IX models, Phys. Rev. D\textbf{82}
    043508 (2010).    
    


\bibitem{awe2} A.~Ashtekar and E.~Wilson-Ewing, {Loop quantum
    cosmology of Bianchi type I models}, Phys. Rev. D\textbf{79} 083535
    (2009).

\bibitem{madrid-bianchi} M.~Martin-Benito, G.~A.~Mena Marugan,
    T.~Pawlowski, Loop quantization of vacuum Bianchi I cosmology,
Phys. Rev. D\textbf{78} 064008 (2008); \\
Physical evolution in loop quantum cosmology: The example of vacuum
Bianchi I, Phys. Rev. D\textbf{80} 084038 (2009). 

\bibitem{awe3} A.~Ashtekar and E.~Wilson-Ewing, Loop quantum
    cosmology of Bianchi type II models, Phys. Rev. D\textbf{80}
    123532 (2009).
    
\bibitem{ck3} A.~Corichi and A.~Karami,
  Loop quantum cosmology of Bianchi IX: inverse triad corrections, unpublished


\bibitem{as} A.~Ashtekar and T.~A.~Schilling, Geometrical
    formulation of quantum mechanics. In: \textit{On
    Einstein's Path: Essays in Honor of Engelbert Sch\"ucking},
     Harvey, A.\ (ed.) (Springer, New York (1999)), 23--65,
     \texttt{arXiv:gr-qc/9706069}



\bibitem{vt} V.~Taveras, LQC corrections to the Friedmann equations
    for a universe with a free scalar field, Phys. Rev.
    \textbf{D78} 064072 (2008).


\bibitem{cv1}  A.~Corichi and T.~Vukasinac,
  Effective constrained polymeric theories and their continuum limit,
  Phys.\ Rev.\ D {\bf 86}, 064019 (2012)
  [arXiv:1202.1846 [gr-qc]].

\bibitem{ps} P.~Singh, Are loop quantum cosmologies never
    singular? Class. Quant. Grav. \textbf{26} 125005 (2009)

\bibitem{cs3} A. Corichi, P. Singh, A geometric perspective on
    singularity resolution and uniqueness in loop quantum cosmology,
    Phys. Rev.D {\bf 80} 044024 (2009)

\bibitem{svv} P.~Singh, K.~Vandersloot, and G.~V. Vereshchagin,
    {Non-singular bouncing universes in loop quantum cosmology},
 Phys. Rev. D{\bf 74} 043510, (2006)

\bibitem{ck-inflation}A.~Corichi and A.~Karami, On
    the measure problem in slow roll inflation and loop quantum
    cosmology, Phys. Rev. D\textbf{83} 104006 (2011)

\bibitem{cs1} A.~Corichi and P.~Singh, Is loop quantization in
    cosmology unique? Phys. Rev. D\textbf{78} 024034 (2008)


\bibitem{gt} G.W. Gibbons, N. Turok, The measure problem in
    cosmology, Phys. Rev. D {\bf 77} 063516 (2008)



\bibitem{ps-fv} P.~Singh and F.~Vidotto,
  Exotic singularities and spatially curved Loop Quantum Cosmology,
  Phys.\ Rev.\ D {\bf 83}, 064027 (2011)
  \texttt{arXiv:1012.1307}

\bibitem{bkl1} V. A. Belinskii, I. M. Khalatnikov and E. M.
    Lifshitz, Oscillatory approach to a singular point in
    the relativistic cosmology, Adv. Phys. \textbf{31} 525-573 (1970)

\bibitem{ahs} A.~Ashtekar, A.~Henderson and D.~Sloan,
    Hamiltonian formulation of General Relativity and the Belinksii,
    Khalatnikov, Lifshitz conjecture, Class. Quant. Grav. \textbf{26}
    052001 (2009);\\
    A Hamiltonian Formulation of the BKL Conjecture, Phys. Rev. D\textbf{83}
    084024 (2011)

\bibitem{CKM} A.~Corichi, A.~Karami and E.~Montoya,
  Loop Quantum Cosmology: Anisotropy and singularity resolution,
  \texttt{1210.7248}.

\bibitem{singh-gupt} B.~Gupt and P.~Singh,
  Quantum gravitational Kasner transitions in Bianchi-I spacetime,
  Phys.\ Rev.\ D {\bf 86}, 024034 (2012)
  \texttt{arXiv:1205.6763}.

\bibitem{ps-bianchi} B.~Gupt and P.~Singh, Contrasting features of
    anisotropic loop quantum cosmologies: the role of spatial
    curvature, Phys. Rev. D\textbf{85}, 044011 (2012).\\
    P.~Singh, Curvature invariants, geodesics and the strength of
    singularities in Bianchi-I loop quantum cosmology,
    \texttt{arXiv:1112.6391}.


\bibitem{ac-bianchi} A.~Corichi and E.~Montoya, Effective dynamics in
    Bianchi type II loop quantum cosmology,
    Phys.\ Rev.\ D {\bf 85}, 104052 (2012)
     \texttt{arXiv:1201.4853}.

\bibitem{puchta} J.~Puchta,  Quantum fluctuations in quantum spacetime. MSc Thesis under supervision of Jerzy Lewandowski, University of Warsaw
 2009. 

\bibitem{dapor1} A.~Dapor and J.~Lewandowski, Emergent Isotropy-Breaking in Quantum Cosmology, arxiv:1211.0161.

\bibitem{dapor2} A.~Dapor, J.~Lewandowski, and  Y.~Tavakoli, Lorentz Symmetry in QFT on Quantum Bianchi I Space-Time, Phys. Rev. D{\bf
    86}  064013 (2012);

 
\bibitem{reportbrandenberger} V. F.~Mukhanov,  H. A.~Feldman, R. H.~Brandenberger, Theory of cosmological perturbations, Phys. Rep. 215, 5-6, 203 (1992).

\bibitem{langlois} D.~Langlois, Hamiltonian formalism and gauge
    invariance for linear perturbations in inflation, Class. Quant.
    Grav. \textbf{11} 389-407 (1994).
    

\bibitem{ghtw} K.~Giesel, S.~Hofmann, T.~Thiemann, O.~Winkler,
    Manifestly gauge-invariant general relativistic perturbation
    theory: I. Foundations, Class. Quant. Grav.
    \textbf{27} 055005 (2010); Manifestly gauge-invariant general relativistic perturbation
    theory: II. FRW background and first order, Class. Quant. Grav.
    \textbf{27} 055006 (2010).

\bibitem{joao2}L.~Bethke and J.~ Magueijo, Inflationary tensor
    fluctuations, as viewed by Ashtekar variables and their
    imaginary friends, Phys. Rev. \textbf{D}84 024014 (2011).
    

\bibitem{dt}B.~Dittrich and J.~Tambornino,
    Gauge invariant perturbations around symmetry reduced sectors of
    General Relativity, Class. Quant. Grav. \textbf{24} 4543-4585
    (2007).

\bibitem{pert_scalar} M. Bojowald, H. H. Hernandez, M. Kagan, P.
    Singh, A. Skirzewski, Hamiltonian cosmological perturbation
    theory with loop quantum gravity corrections, Phys. Rev. D{\bf
    74} 123512 (2006);\\ M. Bojowald, H. H.
    Hernandez, M. Kagan, P. Singh, A. Skirzewski, Formation and
    Evolution of Structure in Loop Cosmology, Phys. Rev. Lett. {\bf
    98} 031301 (2007).


\bibitem{aan3}I.~Agullo, A.~Ashtekar and W.~Nelson, The
    pre-inflationary dynamics of loop quantum cosmology:
    Confronting quantum gravity with observations, 
    \texttt{arXiv:1302.0254}.

\bibitem{wmap} E. Komatsu et al, Seven-year Wilkinson microwave
    anisotropy probe (WMAP) observations: Cosmological
    interpreation, Astrophys. J. Suppl. Ser. \textbf{192}, 18 (2011).


    \bibitem{aan1}I.~Agullo, A.~Ashtekar and W.~Nelson, A quantum
   gravity extension of the inflationary scenario, Phys. Rev.
  Lett.  {\bf 109} 251301 (2012).

    


\bibitem{as3}A.~Ashtekar and D.~Sloan, Probability of inflation in
    loop quantum cosmology, Gen. Rel. Grav. 43, 3619-3656 (2011).
    \texttt{arXiv:1103.2475}


\bibitem{ghs} G. W. Gibbons, S.W. Hawking and J. Stewart, 	A
    Natural Measure on the set of all universes, Nucl.
    Phys. \textbf{B281} 736 (1987)

\bibitem{hp} S.~W.~Hawking and D.~N.~Page, How probable is
    inflation?, Nucl. Phys. \textbf{B298} 789 (1988)

\bibitem{sw} J.~S.~Schiffrin and R.~M.~Wald,
  Measure and Probability in Cosmology,
  Phys.\ Rev.\ D {\bf 86}, 023521 (2012)
  \texttt{arXiv:1202.1818}

\bibitem{klm} L. Kofman, A. Linde,  V. Mukhanov,  Inflationary
    theory and alternative cosmology, JHEP {\bf 0210} 057 (2002)
        
\bibitem{waldbook} R.~M.~Wald, {\em Quantum field theory in curved
    space-time and black hole thermodynamics} (University pf
    Chicago Press, Chicago 1994).        
        
 \bibitem{parker66} L.~Parker, The creation of particles in an
    expanding universe, Ph.D. thesis, Harvard University (1966).


\bibitem{parker69} L.~Parker, Particle creation in expanding
    universes, {\it Phys.Rev.Lett.} {\bf 21} 562 (1968); Quantized
    fields and particle creation in expanding universes 1, {\it
    Phys. Rev.} {\bf 183}, 1057(1969).
    
    
\bibitem{parker-fulling74} L.~Parker and  S.~A.~Fulling, Adiabatic
        regularization of the energy momentum tensor of a quantized
        field in homogeneous spaces, Phys. Rev. D {\bf 9} 341 (1974).

        

\bibitem{holman-tolley} R.~Holman and A.~Tolley, Enhanced Non-Gaussianity from excited states, JCAP \textbf{0805}, 001 (2008).

\bibitem{agullo-parker} I.~Agullo and L.~Parker,  Non-gaussianities
    and the Stimulated creation of quanta in the inflationary
    universe, Phys. Rev. D\textbf{83} 063526 (2011);\\
    Stimulated creation of quanta during inflation and the observable universe
    Gen. Relativ. Gravit. 43, 2541-2545 (2011). 

\bibitem{ganc} J.~Ganc,  Calculating the local-type fNL for slow-roll inflation with a non-vacuum initial state., Phys. Rev. D\textbf{84} 063514 (2011).
    
\bibitem{agullo-navarro-salas-parker} I.~Agullo, J.~Navarro-Salas and L.~Parker,  Enhanced local-type inflationary trispectrum from a non-vacuum initial state, JCAP \textbf{1205}, 019 (2012).
    
 
\bibitem{halo-bias2} I.~Agullo and S.~Shandera, Large non-Gaussian
    halo bias from single field inflation JCAP \textbf{1209}, 007 (2012).
        
        
\bibitem{halo-bias1} J.~Ganc and E.~Komatsu,  Scale dependent bias
    of galaxies and $\mu$-type distprtion of the cosmic microwave
    background spectrum from a single field inflation with a
    modified initial state, Phys. Rev. D \textbf{86} 023518 (2012).

\bibitem{halo-bias3} F.~Schmidt and L.~Hui, CMB power asymmetry from
    Gaussian modulation, Phys. Rev. Lett. {\bf 110} 011301 (2013).

        
\bibitem{penrose-weyl} R. Penrose, Singularities and time-asymmetry,
    in \textit{General Relativity: An Einstein Centenary Survey},
    edited by S. W. Hawking and W. Israel. (Cambridge University Press,
    Cambridge, 1979), pages 581-638.

        
        
        
        
    
    
    
       
\end{thebibliography}
\end{document}